\documentclass[twocolumn]{aastex631}

\usepackage{booktabs}

\usepackage{amsmath}
\newcommand{\sw}[1]{\texttt{#1}}

\usepackage{tikz}

\begin{document}

\title{A Detection of Helium in the Bright Superluminous Supernova SN\,2024rmj}

\author[0000-0003-0871-4641]{Harsh Kumar}
\affiliation{Center for Astrophysics \textbar{} Harvard \& Smithsonian, 60 Garden Street, Cambridge, MA 02138-1516, USA}
\affiliation{The NSF AI Institute for Artificial Intelligence and Fundamental Interactions, USA}

\author[0000-0002-9392-9681]{Edo Berger}
\affiliation{Center for Astrophysics \textbar{} Harvard \& Smithsonian, 60 Garden Street, Cambridge, MA 02138-1516, USA}
\affiliation{The NSF AI Institute for Artificial Intelligence and Fundamental Interactions, USA}

\author[0000-0003-0526-2248]{Peter K.~Blanchard}
\affiliation{Center for Astrophysics \textbar{} Harvard \& Smithsonian, 60 Garden Street, Cambridge, MA 02138-1516, USA}
\affiliation{The NSF AI Institute for Artificial Intelligence and Fundamental Interactions, USA}

\author[0000-0001-6395-6702]{Sebastian Gomez}
\affiliation{Center for Astrophysics \textbar{} Harvard \& Smithsonian, 60 Garden Street, Cambridge, MA 02138-1516, USA}

\author[0000-0002-1125-9187]{Daichi Hiramatsu}
\affiliation{Center for Astrophysics \textbar{} Harvard \& Smithsonian, 60 Garden Street, Cambridge, MA 02138-1516, USA}
\affiliation{The NSF AI Institute for Artificial Intelligence and Fundamental Interactions, USA}

\author[0000-0002-1125-9187]{Alex Gagliano}
\affiliation{Center for Astrophysics \textbar{} Harvard \& Smithsonian, 60 Garden Street, Cambridge, MA 02138-1516, USA}
\affiliation{The NSF AI Institute for Artificial Intelligence and Fundamental Interactions, USA}

\author[0000-0002-1895-6639]{Moira Andrews} 
\affiliation{Las Cumbres Observatory, 6740 Cortona Drive, Suite 102, Goleta, CA 93117-5575, USA}
\affiliation{Department of Physics, University of California, Santa Barbara, CA 93106-9530, USA}

\author{K. Azalee Bostroem}  
\affiliation{Las Cumbres Observatory, 6740 Cortona Drive, Suite 102, Goleta, CA 93117-5575, USA}

\author[0000-0003-4914-5625]{Joseph Farah}   
\affiliation{Las Cumbres Observatory, 6740 Cortona Drive, Suite 102, Goleta, CA 93117-5575, USA} 
\affiliation{Department of Physics, University of California, Santa Barbara, CA 93106-9530, USA}

\author{D. Andrew Howell}   
\affiliation{Las Cumbres Observatory, 6740 Cortona Drive, Suite 102, Goleta, CA 93117-5575, USA} 
\affiliation{Department of Physics, University of California, Santa Barbara, CA 93106-9530, USA}
\author{Curtis McCully} 
\affiliation{Las Cumbres Observatory, 6740 Cortona Drive, Suite 102, Goleta, CA 93117-5575, USA}

\begin{abstract}
We present extensive ultraviolet (UV), optical, and near-infrared (NIR) photometric and spectroscopic observations of the nearby hydrogen-poor superluminous supernova (SLSN-I) SN\,2024rmj at $z=0.1189$.  SN\,2024rmj reached a peak absolute magnitude of $M_g\approx -21.9$, placing it at the luminous end of the SLSN-I distribution. The light curve exhibits a pronounced pre-peak bump ($\approx 60$ d before the main peak) and a post-peak bump ($\approx 55$ d after the main peak). The bulk of the light curve is otherwise well fit by a magnetar spin-down model, with typical values (spin: $\approx 2.1$ ms; magnetic field: $\approx 6 \times 10^{13}$ G; ejecta mass: $\approx 12$ M$_\odot$). The optical spectra exhibit characteristic SLSN-I features and evolution, but with a relatively high velocity of $\approx 8,000$ km s$^{-1}$ post-peak. Most significantly, we find a clear detection of helium in the NIR spectra at \ion{He}{1} $\lambda$1.083$~\mu$m and $\lambda$2.058 $\mu$m, blueshifted by $\approx 15,000$ km s$^{-1}$ ($13$ d before peak) and $\approx 13,000$ km s$^{-1}$ ($40$ d after peak), indicating that helium is confined to the outermost ejecta; based on these NIR detections, we also identify likely contribution from \ion{He}{1} $\lambda$5876 \AA\ in the optical spectra on a similar range of timescales.  This represents the most definitive detection of helium in a bright SLSN-I to date, and indicates that progenitors with a thin helium layer can still explode as SLSNe. 
\end{abstract}


\section{Introduction} 
\label{sec:intro}

Hydrogen-poor superluminous supernovae (SLSNe-I) represent a rare and luminous class of stellar explosions of stripped massive stars, with radiated energies of $\sim 10^{51}$~erg~ (e.g., \citealt{2011Natur.474..487Q, 2011ApJ...743..114C, 2012Sci...337..927G, 2015MNRAS.452.3869N, 2017MNRAS.468.4642I, 2018ApJ...852...81L, 2021A&G....62.5.34N}), comparable to the kinetic energy of normal core-collapse SNe (CCSNe). The large luminosities of SLSNe cannot be explained by the radioactive decay of $^{56}$Ni alone, and instead require an additional energy source. The high luminosity, spectral properties (at early and late time), and diverse light curve properties of SLSN-I, as well as their low metallicity environments, can be explained by the magnetar spin-down model~\citep{2010ApJ...717..245K, 2010ApJ...719L.204W, 2014ApJ...787..138L, 2015MNRAS.454.3311M, 2016ApJ...830...13P, 2017MNRAS.470.3566C,  2017ApJ...850...55N,2024MNRAS.535..471G}.  Alternative models have also been proposed (e.g., fallback accretion on a black hole, hydrogen/helium-poor CSM shell interaction, and pair-instability explosions; e.g., \citealt{2017ApJ...845...85L, 2018ApJ...855....2Q}), but they generally lack expected observational signatures or cannot match the full observed diversity.

SLSNe-I optical spectra are characterized by a blue continuum with \ion{O}{2} absorption features near peak. At later phases, the spectra evolve to resemble those of SNe Ic near their peak~\citep{2010ApJ...724L..16P, 2013ApJ...770..128I, 2018ApJ...855....2Q, 2019ARA&A..57..305G}. While SLSNe-I clearly lack hydrogen, the potential presence of helium has been challenging to evaluate in most events. This is due to the high ejecta velocities, resulting in broad lines, in conjunction with other strong features (e.g., \ion{Na}{1}, \ion{Ca}{2} and \ion{C}{2}) in the vicinity of the prominent optical \ion{He}{1} lines. In a few of the lower luminosity SLSN-I ($M\approx -20$ mag) helium features have been claimed from optical spectra~\citep{2020ApJ...902L...8Y}, but their low luminosity raises the question of whether these events belong to the typical SLSN-I class or are just a luminous version of the normal CCSNe (``luminous SNe''; \citealt{2022ApJ...941..107G}) with significant power from $^{56}$Ni radioactive decay as opposed to a magnetar central engine.

A more robust detection of helium is possible through NIR spectroscopy, via the \ion{He}{1} $\lambda 1.083$ $\mu$m and $\lambda 2.058$ $\mu$m lines. The $1.083$ $\mu$m line may be contaminated by stronger \ion{C}{1} $\lambda 1.069$ $\mu$m and \ion{Mg}{2} $\lambda 1.093$ $\mu$m lines (unless a large helium mass is present or the helium ejecta velocities are higher), but the  $2.058$ $\mu$m line is free of blending and can therefore provide a definite detection of helium, even for modest masses of $\sim 0.1$ M$_\odot$ (e.g., \citealt{2020MNRAS.499..730T}). A key challenge, however, is that only events at $z\lesssim 0.15$ can be observed with coverage of the $2.058$ $\mu$m line from the ground.  Thus, only a handful of SLSNe-I have been observed with NIR spectroscopy to date, and of those only one event (SN\,2019hge, in the ``luminous SN'' range) has shown a likely $2.058$ $\mu$m detection \citep{2020ApJ...902L...8Y}. More recently, \citet{2025arXiv250101485K} used post-peak NIR spectra of the SLSN-I SN\,2024ahr to place an upper limit of $\lesssim 0.05$ M$_\odot$ on the ejecta helium mass.

Here, we report the detection of helium lines in the NIR spectrum of the bright SLSN-I SN\,2024rmj, both before and after peak, making it the first SLSN-I with a definitive helium detection. In \S\ref{sec:obs} we describe the discovery and observations of SN\,2024rmj. We analyze the light curves and spectra, including the detection of helium lines, in \S\ref{sec:analysis}. We discuss the implications of this finding in \S\ref{sec:discussion}, and summarize the key results in \S\ref{sec:conclusion}.

\section{Discovery and Observations}
\label{sec:obs}

\subsection{Discovery} 
\label{sec:discovery}

SN\,2024rmj was discovered by the Zwicky Transient Facility~\citep[ZTF;][]{2019PASP..131a8002B, 2019PASP..131g8001G} survey on 2024 August 4 at 10:38:35 UT at coordinates RA(J2000) = $01^\text{h}$ $07^\text{m}$ $52.74^\text{s}$, Dec(J2000) = $+03^{\circ}$ $30^{\prime}$ $40.3^{\prime\prime}$ (internal identifier name, ZTF24aaysowl; \citealt{2024TNSTR2787....1S}). The SN was independently detected by Pan-STARRS (PS24hqm), ATLAS (ATLAS24mwh), GOTO (GOTO24foa), and MASTER (MASTER OT J010752.76+033039.1).  Spectroscopic follow-up 29 d after discovery revealed broad \ion{O}{2} absorption and a blue continuum, leading to a classification as SLSN-I at a redshift of $z\approx 0.14$ \citep{2024TNSCR3255....1S}. Here we measure a more precise and lower redshift of $z=0.1189\pm 0.0008$ based on higher signal-to-noise ratio spectra that exhibit host galaxy emission lines (see \S\ref{subsec:specobs}).

\subsection{Host Galaxy}
\begin{figure}[t!]
\center
\includegraphics[width=1\linewidth]{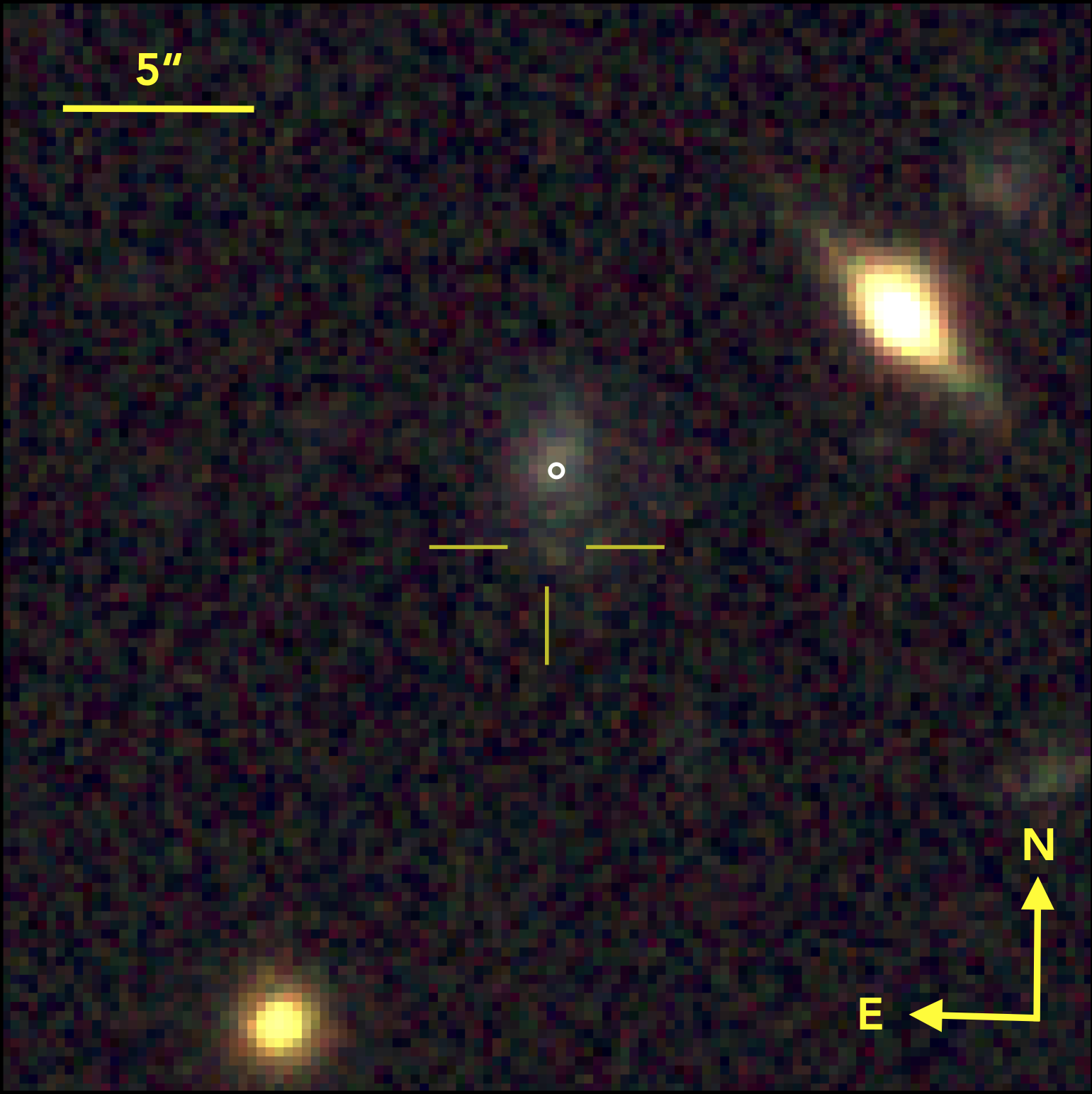}
\caption{The location of SN\,2024rmj (crosshairs) in a pre-explosion Legacy Survey's Data Release 10 image~\citep{2018AAS...23143601F}. The SN is located on the outskirts of its dwarf host galaxy (white circle), $\approx 2.1''\approx 6.8$ kpc from the host center.}
\label{fig:discovery}
\end{figure}

Near the position of SN\,2024rmj we identify a host galaxy visible in SDSS, Pan-STARRS1 $3\pi$, and Legacy Survey images; see Figure~\ref{fig:discovery}. The galaxy centroid is at RA(J2000) = $01^\text{h}$ $07^\text{m}$ $52.72^\text{s}$, Dec(J2000) = $+03^{\circ}$ $30^{\prime}$ $42.44^{\prime\prime}$, indicating that SN\,2024rmj is offset by $\approx 2.1^{\prime \prime}\approx 6.8$ kpc from its host center in Legacy Survey images. From SDSS DR16 data we find host magnitudes of: $m_{u} = 22.14 \pm 0.48$, $m_{g} = 21.70 \pm 0.12$, $m_{r} = 21.39 \pm 0.13$, $m_{i} = 20.98 \pm 0.15$, $m_{z} = 20.70 \pm 0.45$. At the redshift of the galaxy, the corresponding absolute magnitudes are $M_{u} = -16.53 \pm 0.48$, $M_{g} = -16.97 \pm 0.12$, $M_{r} = -17.28 \pm 0.13$, $M_{i} = -17.69 \pm 0.15$ and $M_{z} = -17.97 \pm 0.45$. Using the SDSS luminosity function from \citet{2009MNRAS.399.1106M}, this corresponds to a luminosity of $L_g\approx 6.8\times 10^8 L_{g,\odot} \approx 0.04 L^*_{g}$ and $L_r = 6.1\times 10^8 L_{r,\odot} \approx 0.02 L^*_{r}$, indicating that the host is a dwarf galaxy. The faintness of the host and the significant offset indicate that contamination from host emission in the UV and optical light curves of the SN is negligible.

\subsection{Photometric Observations}

We obtained multi-band photometric observations of SN\,2024rmj using several facilities, including Las Cumbres Observatory ~\citep[LCO;][]{2013PASP..125.1031B}, the Fred Lawrence Whipple Observatory (FLWO\footnote{\url{https://www.cfa.harvard.edu/facilities-technology/cfa-facilities/fred-lawrence-whipple-observatory-mt-hopkins-az}}) 1.2-m telescope equipped with Keplercam\footnote{\url{https://pweb.cfa.harvard.edu/facilities-technology/telescopes-instruments/12-meter-48-inch-telescope}}, and publicly available photometry from ZTF, the Asteroid Terrestrial-impact Last Alert System~\citep[ATLAS;][]{2018PASP..130f4505T}, and the Neil Gehrels {\it Swift} UV/optical Telescope (UVOT; \citealt{2005SSRv..120...95R}). The data presented in this paper span 2024 August 4 to 2025 February 5, when the SN location became Sun-constrained.

We commenced photometric observations with the Sinistro cameras on the 1-m telescopes of the Las Cumbres Observatory (LCO) starting on 2024 September 17 as a part of the Global Supernova Project \citep{2017AAS...23031803H}. We used the $U,B,V,g,r,i$ filters and obtained photometry using point-spread function (PSF) fitting, employing the \texttt{lcogtsnpipe} pipeline~\citep{2016MNRAS.459.3939V}.

We undertook photometric observation with KeplerCam on the 48-inch telescope at FLWO starting on 2024 October 13 in the $g,r,i$ filters. The data were processed using a local Python-based pipeline following standard reduction procedures, and PSF-fit photometry was obtained. 

The {\it Swift} UVOT observations were obtained as a ToO program (PIs: Schulze, Chen) in all six available filters (UVW2, UVM2, UVW1, $U$, $B$, $V$) in 25 epochs spanning from 2024 September 2 to 2025 January 15. We performed aperture photometry using \texttt{UVOTSOURCE}\footnote{\url{https://www.swift.ac.uk/analysis/uvot/mag.php}}, a part of the Swift software suite, with an aperture of $5''$ radius and included standard aperture corrections.

\begin{figure}[t!]
\center
\includegraphics[width=1\linewidth]{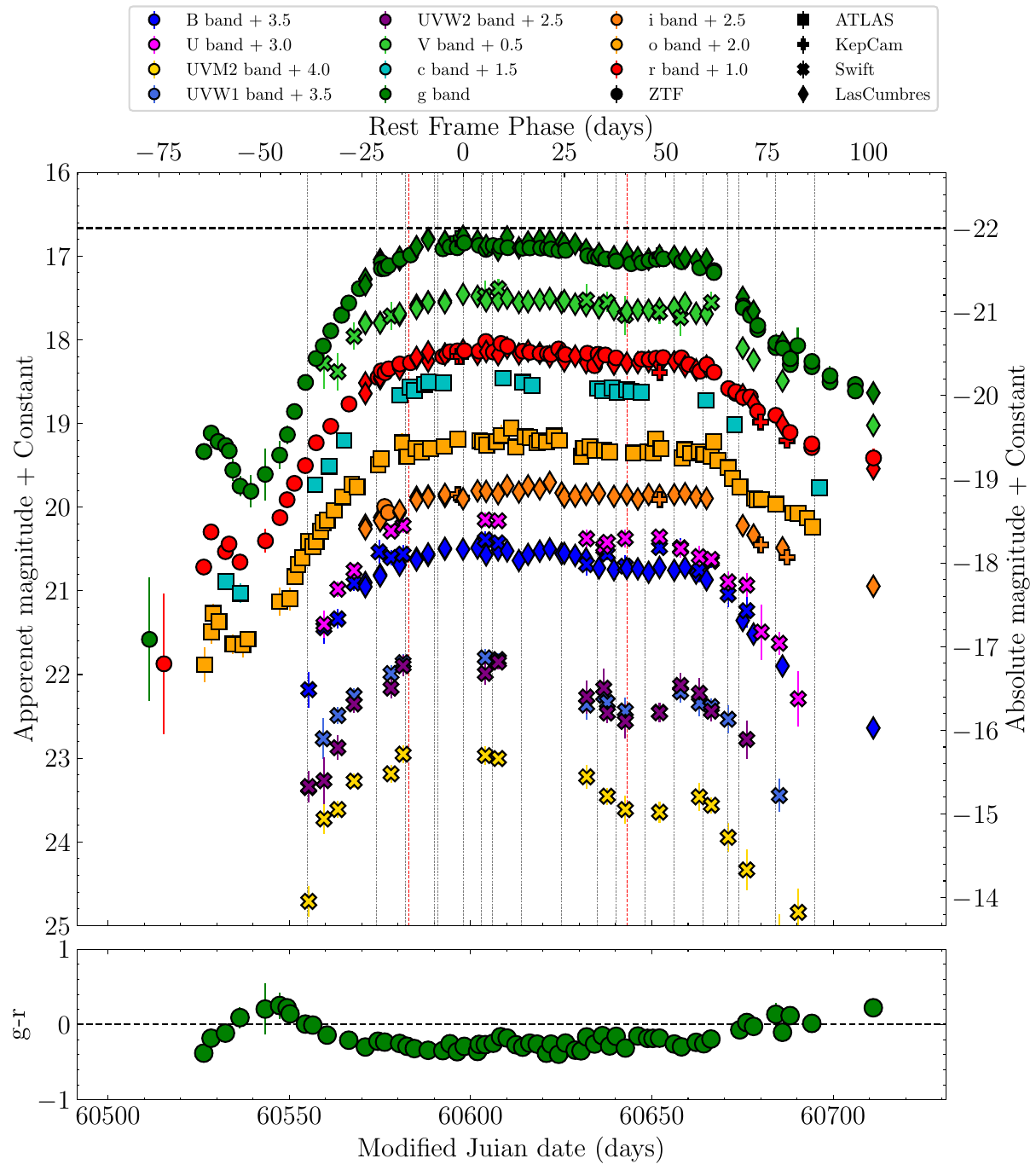}
\caption{Optical/UV light curves of SN\,2024rmj. All magnitudes are in the AB system and are corrected for Galactic Extinction. Vertical lines mark the epochs of optical (gray) and near-IR (red) spectroscopy. SN\,2024rmj rises from discovery to peak in $\approx 70$ d and reaches a peak absolute magnitude of $M_{g}\approx M_r\approx -21.9$  ($K$-corrected). The $g,r,o$-band light curves exhibit a pronounced pre-peak bump, while all filters (particularly the UV) exhibit another bump centered at a phase of $\approx +55$ d. The $g-r$ color (bottom)  reddens during the pre-peak bump, becomes blue again on the rise to the main peak, and then again during the post-peak bump, and eventually reddens on the final decline.}
\label{fig:lc}
\end{figure}
We obtained the ZTF photometry from the Automatic Learning for the Rapid Classification of Events (ALeRCE) broker~\citep{2021AJ....161..242F}. The observations span from 2024 August 4 to 2025 February 5 in the $g,r,i$ filters. Additionally, forced photometry on images prior to the first discovery epoch resulted in two additional photometric data points (S/N $\approx 3$) in $g,~r$ filters.

We obtained ATLAS photometry through the ATLAS Forced Photometry Server~\citep{2021TNSAN...7....1S}. We applied a minimum signal-to-noise ratio of $\geq 5$. The observations span from 2024 August 4 to 2025 January 21 in the $c,o$ filters. 

The resulting photometry is presented in Table~\ref{tab:photometrytable}. The photometric measurements are in the AB magnitude system for all filters, and corrected for Galactic extinction, with $E(B-V) = 0.023$ mag \citep{2011ApJ...737..103S}, assuming the~\cite{1999PASP..111...63F} reddening law with $R_V =3.1$. We use standard Planck18 FlatLambdaCDM cosmology \citep{2020A&A...641A...6P} with Hubble constant $H_{0} = 67.4 \pm 0.5~\mathrm{km~s^{-1}~Mpc^{-1}}$. The resulting luminosity distance is $d_{L} = 573$ Mpc for $z = 0.1189$ (see \S~\ref{subsec:specobs}).

\subsection{Optical Spectroscopic Observations} \label{subsec:specobs}

We utilized the Binospec spectrograph \citep{2019PASP..131g5004F} on the MMT 6.5-m telescope to obtain spectra on 2024 September 29 and October 8. We used the LP3800 filter in combination with 270 lines/mm grating and $1^{\prime \prime}$ wide slit covering a wavelength range of $3825-9200$ \AA\ with a resolution of $\approx 1500$. The data were reduced using the \sw{PypeIt} package~\citep{pypeit:joss_pub} in a standard manner. The one-dimensional spectrum was extracted and flux-calibrated using a standard star observation obtained with the same configuration. 

We used the Low Dispersion Survey Spectrograph 3~\citep[LDSS3;][]{2016ApJ...817..141S} on the Magellan/Clay 6.5-m telescope on 2024 December 27 (during the post-peak bump). We used the VPH-All grism with a 1$\arcsec$ wide slit, covering a wavelength range of $4265 - 9650$ \AA\ with a resolution of $\approx 700$. The data were reduced using the \sw{PypeIt} package. The one-dimensional spectrum was extracted and flux-calibrated using a standard star observation obtained with the same configuration on the same night.  

We used the FLOYDS spectrographs mounted on the 2-m LCO Faulkes telescopes North (FTN) at Haleakalā, USA, and South (FTS) at Siding Spring, Australia, to obtain several spectra from 2024 September 21 to 2025 January 20 as part of the Global Supernova Project. We used a $2\arcsec$ slit placed on the target along the parallactic angle \citep{Filippenko1982}, covering a wavelength range of $3400 - 10000$ \AA\ with a resolution of $\approx 400$. One-dimensional spectra were extracted, reduced, and calibrated following standard procedures using \texttt{floyds\_pipeline}\footnote{\url{https://github.com/LCOGT/floyds_pipeline}} \citep{Valenti2014}. 

The MMT and LDSS spectra exhibit weak H$\alpha$ and [\ion{O}{3}] host galaxy emission lines at a common redshift of $0.1189 \pm 0.0008$, which we use as the redshift of SN\,2024rmj throughout this paper.

\subsection{Near-IR spectroscopic observations}

We used the GNIRS instrument on the Gemini North 8-m telescope on 2024 September 29 and November 28 to obtain NIR spectra (Program ID: GN-2024B-Q-130, PI: Kumar). The spectra cover the wavelength range $0.82- 2.52$ $\mu$m. The data were reduced using the Gemini \sw{Pypeit} package, including flatfielding using GCALflats and sky subtraction obtained on the same night as the Science frames. Flux calibration was performed on the extracted spectra using standard stars. Finally, a telluric correction was applied for the atmospheric absorption features.

\section{Analysis} 
\label{sec:analysis}

\subsection{Photometric Evolution}
\label{subsec:photevo}

The optical and UV light curves of SN\,2024rmj in 11 filters are shown in Figure ~\ref{fig:lc}. The light curve peaks in $g$-band at MJD 60598, which we define as ${\rm phase} = 0$ d.  All timescales below are given in the rest-frame relative to this phase.  The early light curve is marked by an initial rapidly evolving ``bump'' lasting about 20 days, and rising about 3 mag in 10 days to a peak absolute magnitude of $M_g\approx -19.5$ at $-62$ d, before declining by about a magnitude. The light curve then resumes a longer rise over 44 days to a peak absolute magnitude of $M_g\approx -21.9$; the peak mag is in the top quartile of all SLSNe-I \citep{2024MNRAS.535..471G}.  The light curve subsequently declines gradually before rising to a second ``bump'' (most prominent in the UV and blue optical bands) centered at a phase of about $+60$ d and with a duration of about 30 d.  Following the decline of the second bump, the light curve resumes its decline, with a steeper slope in the UV and a somewhat shallower slope in the optical.

The $g-r$ color (Figure~\ref{fig:lc}) rapidly evolves from blue ($\approx -0.4$ mag) to red ($\approx 0.4$ mag) during the pre-peak bump, then evolves to a bluer color of about $-0.3$ mag again during the rise and main peak, beginning to slightly redden on the initial slow decline before becoming slightly bluer again during the post-peak bump, and then rapidly reddening to $\gtrsim 0$ mag after the bump. The color evolution indicates that the pre-peak bump is a separate emission component, akin to post-breakout shock cooling or early circumstellar medium interaction in normal core-collapse SNe, but with a much higher luminosity that may reflect the presence of a powerful central engine \citep{2016MNRAS.457L..79N, 2016ApJ...821...36K}.  Such bumps have been detected in a few previous SLSNe-I \citep{2016MNRAS.457L..79N, 2019MNRAS.487.2215A}, but their origin remains unclear \citep{2016MNRAS.457L..79N}; we defer a discussion of this bump to a separate paper (S.~Gomez et al. in preparation).  The properties of the post-peak bump are similar to those seen in a large fraction of all SLSNe-I~\citep{2022ApJ...933...14H}.

\subsection{Bolometric light curve and  Temperature and Radius evolution}
\label{subsec:bolo}

\begin{figure}[t!]
\center
\includegraphics[width=\linewidth]{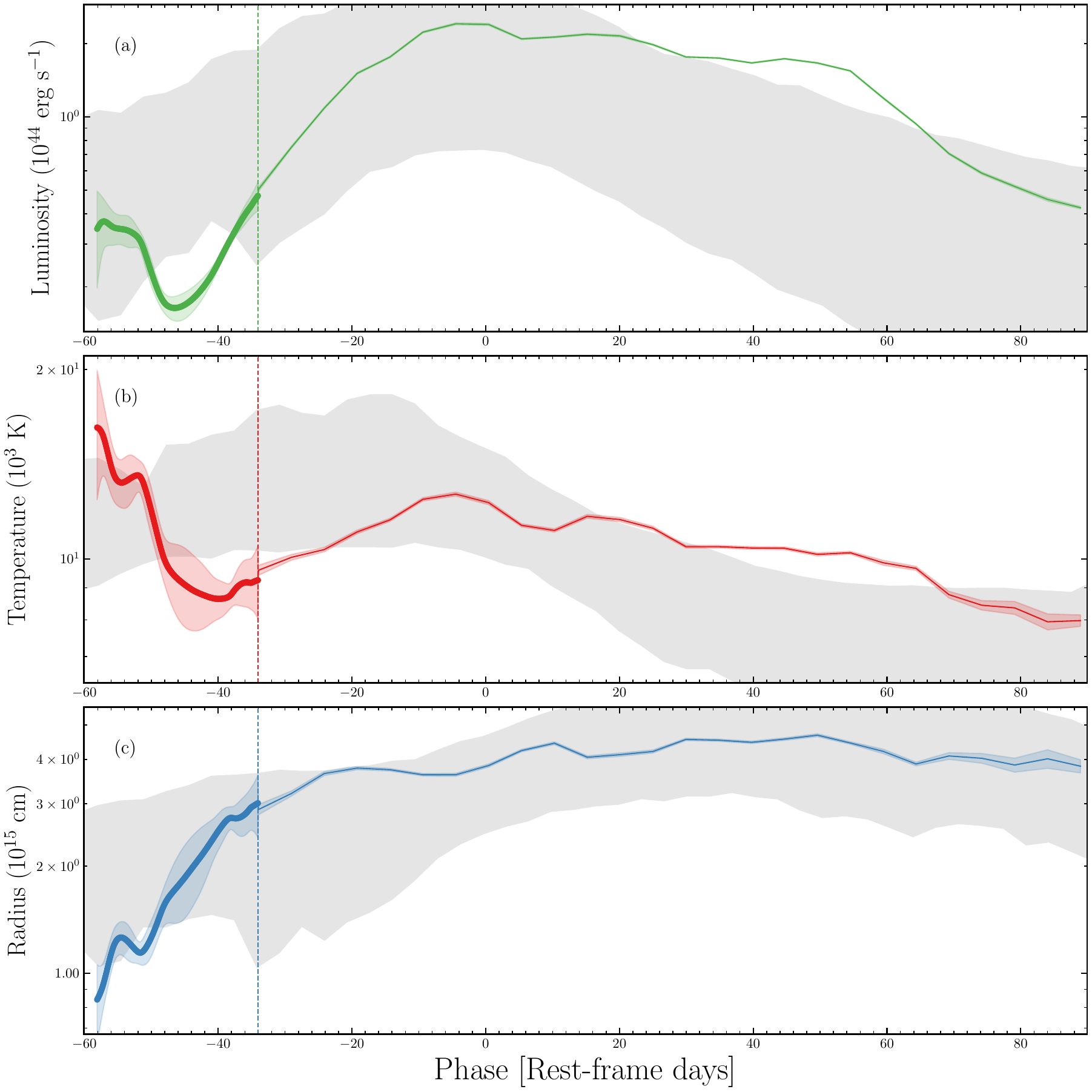} 
\caption{The pseudo-bolometric light curve ($a$), photospheric temperature ($b$), and photospheric radius ($c$) of SN\,2024rmj. The grey bands represent the $1\sigma$ range of parameters of the respective quantities for the SLSN-I sample \citep{2024MNRAS.535..471G}. The early light curve, including the pre-peak bump, for which only $g,r,o$-band data are available is fit separately (left of the vertical dashed lines). We use a bin window of 4 days to reduce fluctuations due to epoch-to-epoch photometric uncertainties. The peak luminosity is $\approx 2.4\times 10^{44}$ erg s$^{-1}$. The pre-peak bump exhibits rapid cooling, while the post-peak bump is manifested mainly as plateauing of the temperature and radius evolution.}
\label{fig:bolometric}
\end{figure}

We use the multi-band photometry to construct a pseudo-bolometric light curve (rest-frame range of $1912-7065$ \AA), and to determine the photospheric temperature and radius evolution, using the \sw{extrabol} package~\citep{2024RNAAS...8...48T}. The results are shown in Figure~\ref{fig:bolometric}.  The luminosity peaks at $\approx 2.4\times10^{44}$ erg s$^{-1}$, typical for the majority of luminous SLSNe-I. The total radiated energy over the observed duration of the light curve is $\approx 1.9\times 10^{51}$ erg, in the top third of the SLSNe-I distribution \citep{2024MNRAS.535..471G}.

The temperature evolution (Figure~\ref{fig:bolometric}) follows the overall $g-r$ color evolution described in \S\ref{subsec:photevo}. It has an initial high value of $\approx 17,000$ K during the pre-peak bump, followed by cooling to $\approx 9,000$ K, and then an increase to $\approx 12,500$ K near peak, followed by a second cooling phase with small fluctuations. The post-peak bump is marked by a flattening of the temperature profile at $\approx 10,000$ K instead of continued cooling as seen in events that lack bumps.  After the bump, the temperature declines to $\approx 7,500$ K.  

The photospheric radius (Figure~\ref{fig:bolometric}) exhibits an initial rapid rise during the pre-peak bump, and then transitions to a more gradual rise during the main peak to $\approx 4.5\times 10^{15}$ cm, where it remains roughly constant before beginning to recede during the decline of the post-peak bump. The long period of steady photospheric radius reflects the impact of the post-peak bump, which is somewhat blended with the initial decline of the light curves. The initial rising phase of the photospheric radius corresponds to a photospheric velocity of $\approx 10,000$ km s$^{-1}$, typical for SLSNe-I.

\begin{figure}[t!]
\center
\includegraphics[width=\linewidth]{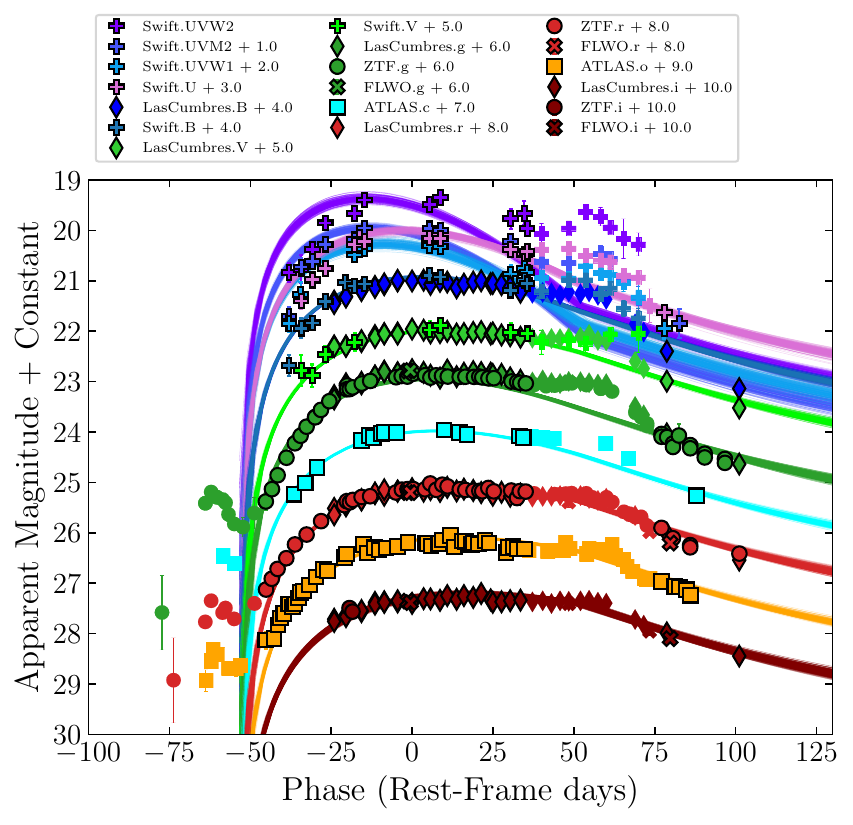} 
\caption{Multi-band MOSFiT model light curves. The fit excludes the pre- and post-peak bumps (data points without a black outline).  The model provides a good fit to the data during the rise, peak, and decline of the main peak, including following the post-peak bump.}
\label{fig:mosfit}
\end{figure}

\begin{table}
\caption{Parameter Posterior Values of the \sw{slsnni} Model from \sw{MOSFiT}.}
\centering
\begin{tabular}{lccc}
\toprule
Parameter &  Prior range & Prior Type & Posterior \\
\midrule
$P_{\mathrm{spin}}$ (ms) & [0.7, 30] & Uniform &$2.14^{+0.33}_{-0.37}$ \\
$B$ ($\times 10^{14}$ G) & log((0,15]) & log  & $-0.20^{+0.19}_{-0.20}$ \\
$M_{\mathrm{ej}}$ (M$_\odot)$ & [0.1, 100] & Uniform &$12.3^{+7.6}_{-2.8}$ \\

$v_{\mathrm{ej}}\,\,(\times 10^3\, {\rm km\,s}^{-1}$) & log([3.5, 5])& log &  $5.0^{+0.4}_{-0.2}$ \\

$t_{\rm exp}\,{\rm (days)}$ & [-200, 0] & Uniform &  $-9.0^{+0.4}_{-0.5}$ \\

$\log\, f_{\rm Ni}$ & log((0, 0.5]) &log & $-2.5^{+0.5}_{-0.4}$ \\

$\log\, n_{\rm H,host}$ & [$10^{16}$, $10^{23}$] &log & $20.6^{+0.1}_{-0.1}$ \\

$\lambda_{\mathrm{cutoff}}$ (\AA) & [2000, 6000] &Uniform &$2985^{+107}_{-137}$ \\

$\alpha$ & [0, 5] & Uniform & $3.65^{+0.52}_{-0.36}$ \\

$T_{\min}\,{\rm (K)}$ & [3000, 1000] & Uniform & $9940^{+50}_{-100}$ \\

$M_{\mathrm{NS}}$ (M$_\odot)$ & $1.7 \pm 0.2$ & Uniform &$1.6^{+0.3}_{-0.4}$ \\

$\theta_{\rm BP}$ &[0, $\pi$/2] & Uniform & $1.0^{+0.4}_{-0.5}$ \\

$\log\, \sigma$ & [$10^{-3}$, $10^2$] &log & $-0.88^{+0.02}_{-0.03}$\\

\bottomrule
\end{tabular}
\label{table:parameters}
\end{table}

\subsection{Magnetar Engine Model}
\label{subsec:model}

\begin{figure*}[t!]
\center\includegraphics[width=0.95\linewidth]{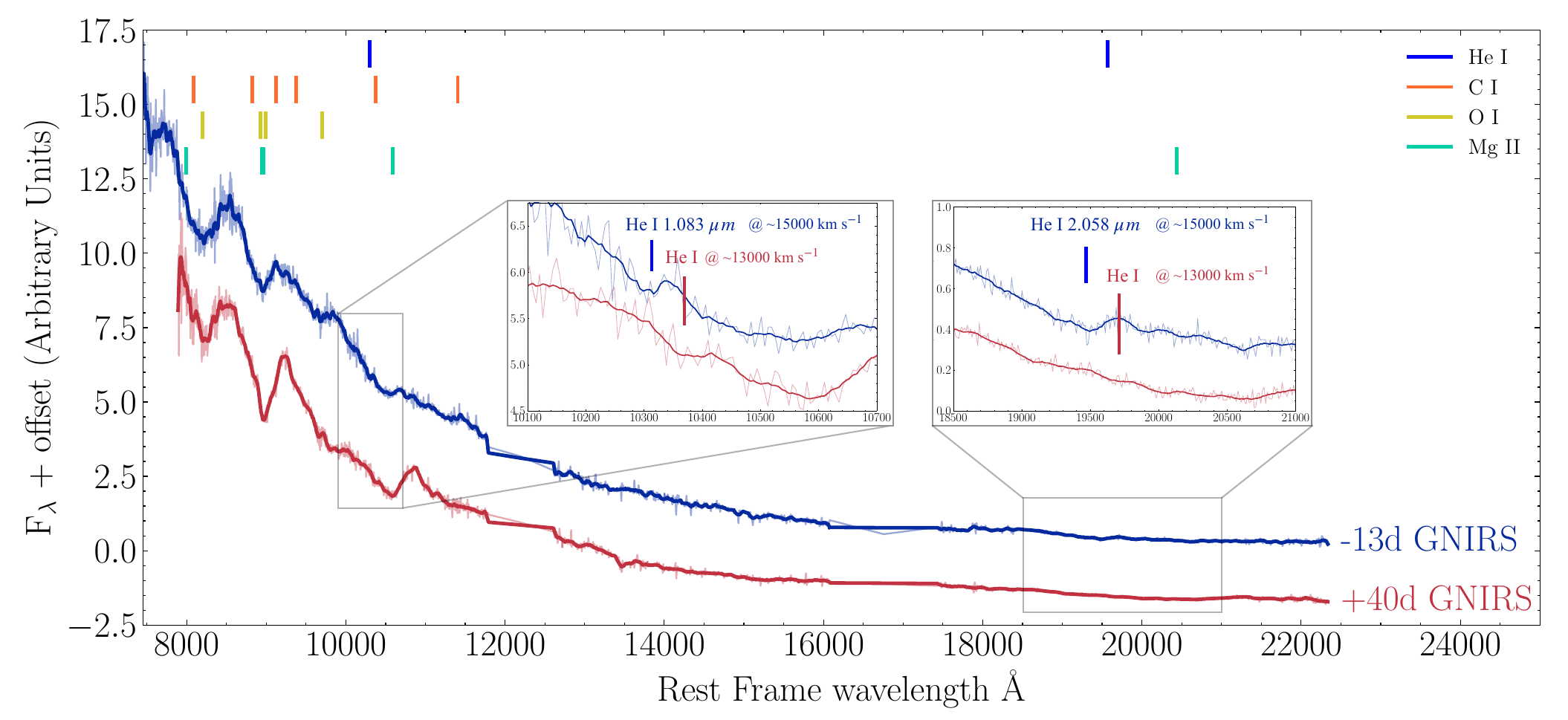}
\caption{Near-IR spectra of SN\,2024rmj obtained at phases of $-13$ d and $+40$ d. The spectrum exhibits broad features of \ion{He}{1}, \ion{C}{1}, and \ion{O}{1} The \ion{He}{1} $\lambda$1.083 $\mu m$ and $\lambda$2.058 $\mu m$ features were detected in the first epoch with a velocity of $-$15000 km s$^{-1}$ (see inset). In the second epoch, the \ion{He}{1} $\lambda$1.083 $\mu m$ and $\lambda$2.058 $\mu m$ (marginally detected) features weaken significantly and shift to the lower velocity of $-$13000 km s$^{-1}$.}
\label{fig:specnir}
\end{figure*}

We model the UV/optical light curve with \sw{MOSFiT} \citep[Modular Open-Source Fitter for Transients;][]{2017ascl.soft10006G, 2018ApJS..236....6G} using the ``\sw{slsnni}'' model extensively used for SLSN-I in previous papers (e.g., \citealt{2017ApJ...850...55N,2024MNRAS.535..471G, 2017ApJ...835L...8N, 2021ApJ...921..180H}). This model combines magnetar spin-down and radioactive decay of $^{56}Ni$ as energy sources. Following the same approach and parameter priors to \citet{2024MNRAS.535..471G}, we generated model light curves and parameter posteriors using data starting from the rest-frame phase of $-51$ d (i.e., after the early-time bump) and excising data at $+30$ to $+68$ d (i.e., the post-peak bump) since the smooth evolution of the model cannot accommodate these features. We employed 150 walkers for 35,000 iterations with a burn-in period of 10,000 steps to ensure the convergence of the model. The best fit model is shown in Figure~\ref{fig:mosfit}, and the parameters are summarized in Table~\ref{table:parameters}. The key parameters are: $P_{\mathrm{spin}}\approx 2.1$ ms, $B\approx 6\times 10^{13}$ G, and $M_{\mathrm{ej}}\approx 12$ M$_{\odot}$, which are typical of the broader SLSN-I population  \citep{2024MNRAS.535..471G}. 

\subsection{NIR Spectroscopic Analysis}

The NIR spectra obtained at phases of $-13$ d and $+40$ d represent some of the most detailed NIR observations of an SLSN-I to date; see Figure~\ref{fig:specnir}. The spectra reveal several prominent features, the most significant being the detection of helium.

\subsubsection{A Detection of Helium}

The \ion{He}{1} $\lambda 2.058$ $\mu$m line is clearly detected in the first, pre-peak NIR spectrum as an absorption feature blueshifted by $\approx 15,000$ km s$^{-1}$, indicating that the helium predominantly lies in the outer, faster-moving ejecta layers. The emission component appears relatively weaker than typically seen in normal SNe Ib, suggesting either a lower amount of helium or less efficient excitation of helium in SN\,2024rmj. 

We also identify the \ion{He}{1} $\lambda 1.083$ $\mu$m feature at the same velocity in the first epoch. In general, this line lies in close proximity to the \ion{C}{1} $\lambda 1.069$ $\mu$m and \ion{Mg}{2} $\lambda 1.091$ $\mu$m lines and is difficult to identify uniquely. However, in the case of  SN\,2024rmj, the higher velocity of helium compared to the photospheric velocity for other elements enables its identification. 

In the second epoch at $+40$d, the helium lines are weaker and have a slightly lower velocity of about $13,000$ km s$^{-1}$, but the $1.083$ $\mu$m line is still distinguishable from the \ion{C}{1} and \ion{Mg}{2} lines.  The detection of both helium lines, in both the pre- and post-peak spectra, provides strong evidence for the presence of helium in the outer ejecta of the SN.

\begin{figure*}[t!]
\center
\includegraphics[width=0.99\textwidth]{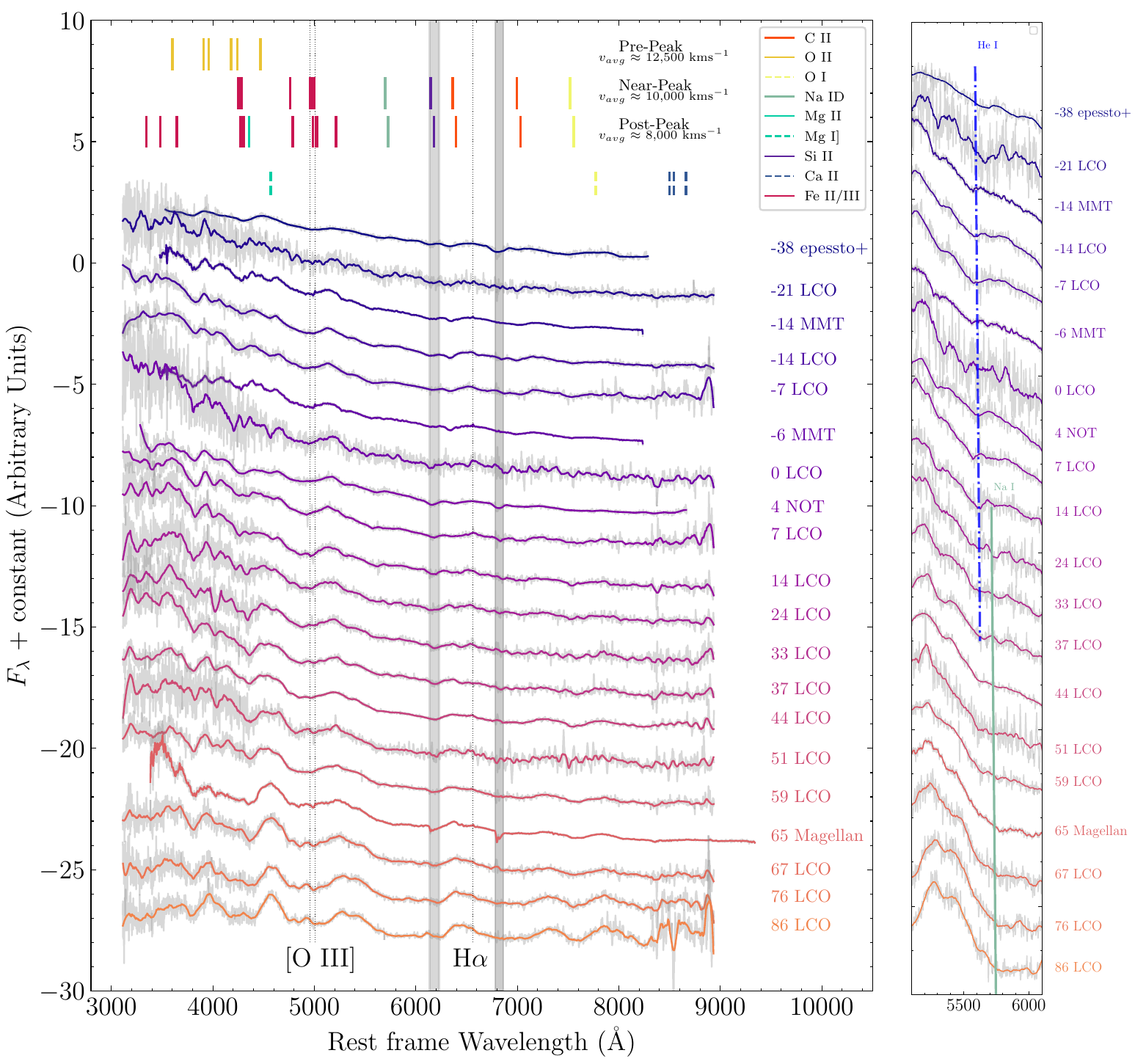}
\caption{{\it Left:} Optical spectra of SN\,2024rmj, spanning phases of $-38$ to $+86$ d. The early time spectra exhibit a blue continuum with the typical \ion{O}{2} lines at a high velocity of $\approx 12,500$ km s$^{-1}$. At peak, features of \ion{Fe}{2}, \ion{Fe}{3}, \ion{C}{2} and \ion{Si}{2} emerge and progressively become stronger.  Emission components of \ion{O}{1}, \ion{Mg}{1}] and the near-IR \ion{Ca}{2} triplet begin to appear in the late spectra. The blueshifted position of features at different phases has been marked.  Vertical grey bands mark the telluric regions.{\it Right:} Zoomed-in view of the region around a likely \ion{He}{1} $\lambda 5876$ \AA\ detection (blue dot-dashed line), at a velocity corresponding to the NIR lines. Later spectra indicate a redward shift as the \ion{Na}{1} D feature (light green line) becomes dominant.}
\label{fig:optical_spectra}
\end{figure*}

\subsubsection{Other NIR Features}

In addition to the \ion{He}{1} features, we identify several other spectral features of \ion{O}{1}, \ion{C}{1}, and \ion{Mg}{2} in the $0.8 - 1.1$ $\mu m$ region of the NIR band. The first broad feature at $\approx 0.82$ $\mu m$ is a blend of strong \ion{O}{1} $\lambda 0.8446$ $\mu m$, relatively weaker \ion{Mg}{2} $\lambda 0.8235$ and \ion{C}{1} $\lambda 0.8833$ $\mu m$ features. The second broad feature at $\approx0.9$ $\mu m$ is a blend of \ion{O}{1} $\lambda 0.9405$ $\mu m$, \ion{C}{1} $\lambda 0.9202$, $\lambda 0.9266$ $\mu m$ and \ion{Mg}{2} $\lambda 0.9218$,  $\lambda 0.9244$ $\mu m$ features. The third and final broad feature in the $1.05 \mu m$ region is a blend of \ion{Mg}{2} $\lambda 1.0914$ $\mu$m and \ion{C}{1} $\lambda 1.0691$ $\mu$m features. The $1.05$ $\mu m$ feature strengthens in the $+$40 d spectrum significantly, likely due to the enhanced \ion{Mg}{2} line contribution. A weak \ion{Mg}{2} $\lambda 2.1062$ $\mu$m line is also detected in the second epoch, confirming the presence of \ion{Mg}{2}. We find velocities of $\approx 11,000$ km s$^{-1}$ and $\approx 9,000$ km s$^{-1}$ in the first and second spectra, respectively. These features are commonly seen in SNe Ib/c at similar epochs \citep{2022ApJ...925..175S}.

\subsection{Optical Spectroscopic Evolution} 
\label{subsec:spec}

The optical spectral sequence, covering phases of $\approx -34$ to $+80$ d, exhibits a characteristic feature evolution seen in SLSNe-I; see Figure~\ref{fig:optical_spectra}. In the pre-peak phase, the spectra display a blue continuum with a blackbody temperature of $\gtrsim 13,000$ K. The spectra also exhibit a series of \ion{O}{2} lines, including the characteristic ``W''-shaped absorption complexes in the $3500-4500$ \AA~range and marginal \ion{Fe}{2} and \ion{Fe}{3} absorption features. Starting around $-19$ d, the broad and blended \ion{Fe}{2} and \ion{Fe}{3} absorption features are clearly visible, which become more pronounced near peak. Using the \ion{O}{2} features at $\approx 3800$ \AA~and $\approx 4000$ \AA, we estimate photospheric expansion velocities of  $\approx 15,000$~km s$^{-1}$ at $-32$ d and $\approx 11,000$ km s$^{-1}$ at 0 d, which is in agreement with the velocity range observed in other SLSNe-I during the pre-peak phase \citep{2025arXiv250321874A}.

Approaching the light curve peak, the \ion{O}{2} features begin to weaken and eventually disappear at $\approx +30$ d. 
Concurrently, the spectra begin to exhibit the \ion{Fe}{2} and \ion{Mg}{2} absorption features at $\approx 4000-5000$ \AA~and the \ion{Si}{2} and \ion{C}{2} features at $\approx 6000-7000$ \AA. The \ion{O}{1} absorption feature emerges near peak phase and becomes stronger at later times. The \ion{Fe}{2} $\lambda 5169$ \AA\ feature commonly used to determine photospheric velocity at these phases is blended in the \ion{Fe}{2} and \ion{Fe}{3} complex, and we therefore use the more isolated \ion{Si}{2} doublet $\lambda \lambda  6347, 6371$ \AA\ features to estimate a velocity of $\approx 11,000$ ~km s$^{-1}$ near peak, again typical of the SLSNe-I population \citep{2025arXiv250321874A}. 

In the post-peak phase, the continuum becomes progressively cooler with temperature dropping below $\approx 10,000$ K. The spectra resemble near peak spectra of SNe Ic/Ic-BL, dominated by features of \ion{Fe}{2}, \ion{Na}{1}D $\lambda\lambda$5890,5896, and \ion{O}{1} $\lambda$7774 absorption features. At $+71$d, the \ion{O}{1} $\lambda$7774 \AA\ feature transforms from absorption to emission feature, and the weak \ion{Ca}{2} NIR triplet ($\lambda \lambda \lambda$ 8498, 8542, 8662) starts to appear. The velocities are $\approx 8,000$~km s$^{-1}$, somewhat higher than in most SLSNe. The photospheric velocities continue to decrease, stabilizing around $\approx 7,500$~km s$^{-1}$. We find no significant spectral evolution during the post-peak bump phase.

\section{Discussion} 
\label{sec:discussion}

\subsection{Possible helium Detection in the Optical Spectra}

There are several \ion{He}{1} lines in the optical range, with the lines at $5876$ \AA\ and $3888$ \AA\ being the most prominent. However, as discussed earlier, it is difficult to confidently identify these lines in SLSNe-I spectra due to blending with stronger broad lines of \ion{Ca}{1} and \ion{Na}{1} D in their vicinity \citep{2016ApJ...826...39N, 2016MNRAS.458.3455M,2019ApJ...882..102G}. In a few events (all being ``luminous SNe'', $M_r\approx -20$), particularly strong \ion{He}{1} detections in optical spectra have been claimed ~\citep{2020ApJ...902L...8Y}. 

The detection of helium in our NIR spectra raises the question of whether subtle helium lines are discernible in the optical spectra. We use the blueshifted velocity of $\approx 15,000$ km s$^{-1}$ inferred from the NIR lines to search for the strongest \ion{He}{1} line at $5876$ \AA. We identify a likely feature, and find that due to the high velocity of the \ion{He}{1} $5876$ \AA\ line, it is clearly resolved from the nearby \ion{Na}{1}D feature, which has a lower velocity of $\approx 10,000$ km s$^{-1}$. Post-peak the helium feature weakens, while \ion{Na}{1}D becomes stronger and we observe a clear shift to lower velocity, and a blend with the overall broad absorption feature sequence of \ion{Na}{1}, \ion{Si}{2} and \ion{C}{2} in the 5400-6200 \AA\ range (Figure~\ref{fig:optical_spectra}).

The presence of blueshifted \ion{He}{1} $\lambda 5876$ \AA\ in our optical spectra suggests that a careful analysis of this region of the spectrum may be a viable path to additional helium detections in the bulk of the SLSNe-I population using existing optical spectra \citep{2025arXiv250321874A}.

\subsection{Helium Excitation Mechanisms}

The observed helium features require an efficient excitation mechanism. In normal SNe Ib, the excitation is attributed to non-thermal electrons produced by the radioactive decay of $^{56}$Ni in the ejecta layers that contain helium \citep{1991ApJ...383..308L, 2012MNRAS.422...70H}.  However, in SLSNe-I, the blue spectra, even in post-peak phases, suggest that mixing of $^{56}$Ni into the outer ejecta layers may not be efficient, implying that helium excitation requires an alternative source. 

The magnetar powering the SLSN can provide an alternative excitation mechanism. In the magnetar scenario, the energy deposition is more centralized but can affect the entire ejecta through a cascade of processes \citep{2010ApJ...717..245K}. High-energy photons from the magnetar can Compton scatter on electrons, producing a population of non-thermal electrons that can efficiently excite helium atoms through collisional processes \citep{2020ApJ...902L...8Y, 2024A&A...692A.204D}. The weakening of the helium lines between the pre- and post-peak spectra may support this central engine driven excitation as the ejecta layers expand.  This is different from normal SNe Ib in which mixed $^{56}$Ni continues to provide an efficient excitation mechanism even at late phases.

\section{Conclusions} 
\label{sec:conclusion}

We presented detailed UV, optical, and NIR photometric and spectroscopic observations of SN\,2024rmj at $z=0.1189$, with a focus on a search for helium using NIR spectra. The key findings are as follows: 

\begin{itemize}
    \item The SN was discovered early and exhibits a pronounced pre-peak bump, previously detected in only a few SLSNe-I. The light curve then rises on a timescale of $\approx 50$ d to a peak brightness of $M_g\approx -21.9$ mag, (pseudo-bolometric luminosity of $\approx 2.4 \times 10^{44}$ erg s$^{-1}$), which is above median for the SLSN-I population. Following the peak, the light curves exhibit a second bump centered at a phase of $\approx +55$ d and with a width of $\approx 25$ d.

    \item The main light curve is well described by a magnetar spin-down model, with parameters typical of the SLSN-I population; there is no evidence for significant contribution from radioactive decay of $^{56}$Ni.
    
    \item We identify the \ion{He}{1} 1.083 and 2.058 $\mu$m lines in the NIR spectra at high blueshift velocities of $\approx 15,000$ and $\approx 13,000$ km s$^{-1}$, in the pre- and post-peak spectra, respectively.  These are the first confident helium detections in a bright SLSN-I.
    
    \item We identify several other NIR spectral features from \ion{O}{1}, \ion{C}{1}, and \ion{Mg}{2}, which evolve from higher to lower velocity between the two epochs. These features are commonly observed in SNe Ib/c \citep{2022ApJ...925..175S}, as well as  SLSNe-I that have NIR spectral coverage~\citep{2016ApJ...826...39N, 2023ApJ...951...34T, 2017ApJ...840...57Y}.
    
    \item The optical spectra exhibit the typical features and evolution of the SLSN-I population. However, the spectral features at late phases, $\gtrsim +60$ d, are relatively broad compared to other SLSNe-I, indicative of higher ejecta velocities. Additionally, the spectra exhibit a likely weak \ion{He}{1} $\lambda$5876 \AA\ feature until $\approx \approx +35$ d.
    
    \item The presence of \ion{He}{1} suggests that the progenitor of SN\,2024rmj retained some amount of helium prior to the explosion. However, the helium features are significantly weaker than in normal SNe Ib. It is unclear if the amount of helium retained is substantially lower, the excitation mechanism of helium is less efficient, or both.
    
    \item SN\,2024rmj represents the most definitive detection of helium in a bright SLSN-I to date, thanks to its brightness and the high-quality NIR spectroscopic data.
    
\end{itemize}

The detection of helium in SN\,2024rmj highlights the importance of systematic NIR spectroscopy of SLSNe-I, as a way of constraining the composition and state of the progenitor before explosion.  This can be performed from ground-based telescopes for SLSNe-I at $z\lesssim 0.15$, but a broader exploration (potentially testing for redshift evolution) requires NIR spectroscopy with {\it JWST}.  Quantifying the abundance of helium and the excitation mechanism will require detailed spectral modeling targeted at magnetar-powered explosions.

\begin{acknowledgments}
We thank Kali Salmas, Alejandra Milone, and  Benjamin Weiner for scheduling the MMT Binospec observations and Yuri Beletsky for performing the Magellan LDSS-3 observations.

The Berger Time-Domain research group at Harvard is supported by the NSF and NASA grants. The LCO supernova group is supported by NSF grants AST-1911151 and AST-1911225. This work is supported by the National Science Foundation under Cooperative Agreement PHY-2019786 (The NSF AI Institute for Artificial Intelligence and Fundamental Interactions, http://iaifi.org/)

This paper includes data gathered with the 6.5-meter Magellan Telescopes located at Las Campanas Observatory, Chile.

Observations reported here were obtained at the MMT Observatory, a joint facility of the Smithsonian Institution and the University of Arizona. This paper uses data products produced by the OIR Telescope Data Center, supported by the Smithsonian Astrophysical Observatory.

This work makes use of observations from the Las Cumbres Observatory global telescope network. The authors wish to recognize and acknowledge the very significant cultural role and reverence that the summit of Haleakalā has always had within the indigenous Hawaiian community. We are most fortunate to have the opportunity to conduct observations from the mountain. 

We acknowledge the use of public data from the Swift data archive.

This research made use of \sw{PypeIt},\footnote{\url{https://pypeit.readthedocs.io/en/latest/}}
a Python package for semi-automated reduction of astronomical slit-based spectroscopy
\citep{pypeit:joss_pub, pypeit:zenodo}. This research made use of WISeREP\footnote{\url{ URL https://wiserep.org }}~\citep{2012PASP..124..668Y}.

This work has made use of data from the Zwicky Transient Facility (ZTF). ZTF is supported by NSF grant No. AST- 1440341 and a collaboration including Caltech, IPAC, the Weizmann Institute for Science, the Oskar Klein Center at Stockholm University, the University of Maryland, the University of Washington, Deutsches Elektronen-Synchrotron and Humboldt University, Los Alamos National Laboratories, the TANGO Consortium of Taiwan, the University of Wisconsin–Milwaukee, and Lawrence Berkeley National Laboratories. Operations are conducted by COO, IPAC, and UW. The ZTF forced-photometry service was funded under the Heising-Simons Foundation grant No. 12540303 (PI: Graham).

This work has made use of data from the Asteroid Terrestrial-impact Last Alert System (ATLAS) project. The Asteroid Terrestrial-impact Last Alert System (ATLAS) project is primarily funded to search for near-earth asteroids through NASA grants NN12AR55G, 80NSSC18K0284, and 80NSSC18K1575; byproducts of the NEO search include images and catalogs from the survey area. This work was partially funded by Kepler/K2 grant J1944/80NSSC19K0112 and HST GO-15889, and STFC grants ST/T000198/1 and ST/S006109/1. The ATLAS science products have been made possible through the contributions of the University of Hawaii Institute for Astronomy, the Queen's University Belfast, the Space Telescope Science Institute, the South African Astronomical Observatory, and The Millennium Institute of Astrophysics (MAS), Chile.

This research has made use of the NASA Astrophysics Data System (ADS), the NASA/IPAC Extragalactic Database (NED), and NASA/IPAC Infrared Science Archive (IRSA, which is funded by NASA and operated by the California Institute of Technology) and IRAF (which is distributed by the National Optical Astronomy Observatory, NOAO, operated by the Association of Universities for Research in Astronomy, AURA, Inc., under cooperative agreement with the NSF).

TNS is supported by funding from the Weizmann Institute of Science, as well as grants from the Israeli Institute for Advanced Studies and the European Union via ERC grant No. 725161.

\end{acknowledgments}

\vspace{5mm}
\facilities{ATLAS, FLWO 1.2m, LCO, Magellan, MMT, Swift(UVOT), and ZTF}

\software{astropy~\citep{2013A&A...558A..33A,2018AJ....156..123A, 2022ApJ...935..167A}, \sw{SExtractor}~\citep{1996A&AS..117..393B} \sw{NumPy}~\citep{harris2020array}, \sw{photutils}~\citep{2022zndo...6825092B}, \sw{PyRAF}~\citep{2012ascl.soft07011S}, \sw{SciPy}~\citep{2020SciPy-NMeth}, \sw{UVOTSOURCE}, \sw{extrabol}~\citep{2024RNAAS...8...48T}, \sw{PypeIt}~\citep{pypeit:joss_pub}, \sw{lcogtsnpipe}~\citep{2016MNRAS.459.3939V}, and \sw{MOSFiT}~\citep{2017ascl.soft10006G},  }
\newpage

\appendix
\section{Photometry}

\begin{longtable}{|c|c|c|c|} 
\hline
\hline
MJD & Filter & Magnitude $\pm$  e\_magnitude & Telescope \\
\hline
\hline
60511.42786	& g & $ 	21.58	\pm	0.74	$	& ZTF	\\
60515.42264	&	r & $ 	20.87	\pm	0.84	$	& ZTF	\\
60526.40024	&	r & $ 	19.72	\pm	0.08	$	& ZTF	\\
60526.44346	&	g & $ 	19.34	\pm	0.09	$	& ZTF	\\
60526.44346	&	g & $ 	19.34	\pm	0.09	$	& ZTF	\\
60526.59826	&	o & $ 	19.88	\pm	0.21	$	& ATLAS	\\
60528.4234	&	r & $ 	19.3	\pm	0.06	$	& ZTF	\\
60528.44332	&	g & $ 	19.11	\pm	0.1	$	& ZTF	\\
60528.49819	&	o & $ 	19.49	\pm	0.15	$	& ATLAS	\\
60529.06652	&	o & $ 	19.27	\pm	0.11	$	& ATLAS	\\
60530.38329	&	g & $ 	19.22	\pm	0.1	$	& ZTF	\\
60530.38329	&	g & $ 	19.22	\pm	0.1	$	& ZTF	\\
60530.56432	&	o & $ 	19.37	\pm	0.09	$	& ATLAS	\\
60532.37995	&	r & $ 	19.53	\pm	0.05	$	& ZTF	\\
60532.42736	&	g & $ 	19.26	\pm	0.04	$	& ZTF	\\
60532.51653	&	c & $ 	19.39	\pm	0.09	$	& ATLAS	\\
60533.36975	&	r & $ 	19.44	\pm	0.04	$	& ZTF	\\
60533.42683	&	g & $ 	19.33	\pm	0.04	$	& ZTF	\\
60534.42294	&	g & $ 	19.56	\pm	0.13	$	& ZTF	\\
60534.48549	&	o & $ 	19.64	\pm	0.12	$	& ATLAS	\\
60535.33153	&	c & $ >	19.29			$	& ATLAS	\\
60536.39544	&	r & $ 	19.66	\pm	0.07	$	& ZTF	\\
60536.42302	&	g & $ 	19.75	\pm	0.13	$	& ZTF	\\
60536.54232	&	c & $ 	19.53	\pm	0.11	$	& ATLAS	\\
60537.1378	&	o & $ 	19.64	\pm	0.15	$	& ATLAS	\\
60538.56821	&	o & $ 	19.58	\pm	0.08	$	& ATLAS	\\
60539.41024	&	g & $ 	19.81	\pm	0.19	$	& ZTF	\\
60540.50927	&	o & $ >	18.73			$	& ATLAS	\\
60541.01432	&	o & $ >	18.39			$	& ATLAS	\\
60543.35404	&	g & $ 	19.61	\pm	0.31	$	& ZTF	\\
60543.4265	&	r & $ 	19.4	\pm	0.15	$	& ZTF	\\
60547.32758	&	o & $ 	19.13	\pm	0.17	$	& ATLAS	\\
60547.34252	&	g & $ 	19.38	\pm	0.17	$	& ZTF	\\
60547.42002	&	r & $ 	19.13	\pm	0.06	$	& ZTF	\\
60549.39024	&	r & $ 	18.91	\pm	0.04	$	& ZTF	\\
60549.46325	&	g & $ 	19.13	\pm	0.11	$	& ZTF	\\
60550.20176	&	o & $ 	19.1	\pm	0.14	$	& ATLAS	\\
60551.39979	&	r & $ 	18.72	\pm	0.03	$	& ZTF	\\
60551.42149	&	g & $ 	18.86	\pm	0.07	$	& ZTF	\\
60551.51087	&	o & $ 	18.83	\pm	0.06	$	& ATLAS	\\
60552.38583	&	o & $ 	18.69	\pm	0.06	$	& ATLAS	\\
60553.56824	&	o & $ 	18.6	\pm	0.05	$	& ATLAS	\\
60554.4211	&	r & $ 	18.5	\pm	0.03	$	& ZTF	\\
60554.47701	&	g & $ 	18.51	\pm	0.07	$	& ZTF	\\
60555.27302	&	o & $ 	18.41	\pm	0.05	$	& ATLAS	\\
60555.34515	&	UVW1 & $ 	19.85	\pm	0.16	$	& Swift	\\
60555.34721	&	B & $ 	18.68	\pm	0.22	$	& Swift	\\
60555.35215	&	UVW2 & $ 	20.84	\pm	0.19	$	& Swift	\\
60555.35318	&	V & $ >	17.88			$	& Swift	\\
60555.36112	&	UVM2 & $ 	20.71	\pm	0.19	$	& Swift	\\
60556.48056	&	o & $ 	18.47	\pm	0.05	$	& ATLAS	\\
60557.12348	&	c & $ 	18.23	\pm	0.06	$	& ATLAS	\\
60557.36743	&	o & $ 	18.41	\pm	0.03	$	& ATLAS	\\
60557.38035	&	g & $ 	18.22	\pm	0.08	$	& ZTF	\\
60557.42237	&	r & $ 	18.23	\pm	0.02	$	& ZTF	\\
60558.60758	&	o & $ 	18.29	\pm	0.03	$	& ATLAS	\\
60559.25647	&	o & $ 	18.18	\pm	0.04	$	& ATLAS	\\
60559.35834	&	UVW1 & $ 	19.27	\pm	0.24	$	& Swift	\\
60559.36175	&	g & $ 	18.07	\pm	0.06	$	& ZTF	\\
60559.55511	&	U & $ 	18.4	\pm	0.16	$	& Swift	\\
60559.55561	&	B & $ 	17.94	\pm	0.19	$	& Swift	\\
60559.55786	&	UVW2 & $ 	20.77	\pm	0.28	$	& Swift	\\
60559.55835	&	V & $ 	17.78	\pm	0.31	$	& Swift	\\
60559.56112	&	UVM2 & $ 	19.73	\pm	0.18	$	& Swift	\\
60560.42351	&	o & $ 	18.16	\pm	0.04	$	& ATLAS	\\
60560.96511	&	c & $ 	18.01	\pm	0.03	$	& ATLAS	\\
60561.43627	&	g & $ 	17.89	\pm	0.06	$	& ZTF	\\
60561.44154	&	r & $ 	18.03	\pm	0.03	$	& ZTF	\\
60562.42801	&	o & $ 	18.04	\pm	0.03	$	& ATLAS	\\
60563.41177	&	UVW1 & $ 	18.99	\pm	0.1	$	& Swift	\\
60563.4128	&	U & $ 	17.98	\pm	0.09	$	& Swift	\\
60563.41383	&	B & $ 	17.83	\pm	0.12	$	& Swift	\\
60563.41883	&	UVW2 & $ 	20.37	\pm	0.15	$	& Swift	\\
60563.41986	&	V & $ 	17.88	\pm	0.22	$	& Swift	\\
60563.42709	&	UVM2 & $ 	19.61	\pm	0.1	$	& Swift	\\
60564.41464	&	g & $ 	17.7	\pm	0.02	$	& ZTF	\\
60564.41464	&	g & $ 	17.71	\pm	0.05	$	& ZTF	\\
60564.70458	&	o & $ 	17.88	\pm	0.02	$	& ATLAS	\\
60565.08012	&	c & $ 	17.7	\pm	0.03	$	& ATLAS	\\
60566.40149	&	g & $ 	17.56	\pm	0.05	$	& ZTF	\\
60566.46304	&	r & $ 	17.77	\pm	0.02	$	& ZTF	\\
60567.22861	&	o & $ 	17.73	\pm	0.03	$	& ATLAS	\\
60567.87291	&	UVW1 & $ 	18.75	\pm	0.09	$	& Swift	\\
60567.87392	&	U & $ 	17.76	\pm	0.08	$	& Swift	\\
60567.87494	&	B & $ 	17.41	\pm	0.09	$	& Swift	\\
60567.87982	&	UVW2 & $ 	19.85	\pm	0.11	$	& Swift	\\
60567.88083	&	V & $ 	17.46	\pm	0.16	$	& Swift	\\
60567.88821	&	UVM2 & $ 	19.27	\pm	0.09	$	& Swift	\\
60568.66752	&	o & $ 	17.76	\pm	0.04	$	& ATLAS	\\
60569.31007	&	g & $ 	17.39	\pm	0.06	$	& ZTF	\\
60570.97488	&	B & $ 	17.4	\pm	0.05	$	& LasCumbres	\\
60570.97656	&	B & $ 	17.45	\pm	0.05	$	& LasCumbres	\\
60570.97842	&	V & $ 	17.29	\pm	0.06	$	& LasCumbres	\\
60570.97977	&	V & $ 	17.3	\pm	0.05	$	& LasCumbres	\\
60570.98128	&	g & $ 	17.27	\pm	0.04	$	& LasCumbres	\\
60570.98298	&	g & $ 	17.34	\pm	0.04	$	& LasCumbres	\\
60570.98484	&	r & $ 	17.52	\pm	0.05	$	& LasCumbres	\\
60570.98619	&	r & $ 	17.64	\pm	0.06	$	& LasCumbres	\\
60570.98771	&	i & $ 	17.71	\pm	0.07	$	& LasCumbres	\\
60570.98905	&	i & $ 	17.76	\pm	0.07	$	& LasCumbres	\\
60574.44922	&	r & $ 	17.45	\pm	0.06	$	& ZTF	\\
60574.52301	&	o & $ 	17.5	\pm	0.03	$	& ATLAS	\\
60574.76001	&	B & $ 	17.03	\pm	0.14	$	& Swift	\\
60575.05536	&	B & $ 	17.31	\pm	0.03	$	& LasCumbres	\\
60575.05705	&	B & $ 	17.32	\pm	0.03	$	& LasCumbres	\\
60575.05891	&	V & $ 	17.29	\pm	0.03	$	& LasCumbres	\\
60575.06026	&	V & $ 	17.3	\pm	0.02	$	& LasCumbres	\\
60575.06178	&	g & $ 	17.04	\pm	0.01	$	& LasCumbres	\\
60575.06347	&	g & $ 	17.07	\pm	0.01	$	& LasCumbres	\\
60575.06536	&	r & $ 	17.47	\pm	0.02	$	& LasCumbres	\\
60575.06671	&	r & $ 	17.44	\pm	0.02	$	& LasCumbres	\\
60575.06823	&	i & $ 	17.62	\pm	0.02	$	& LasCumbres	\\
60575.06958	&	i & $ 	17.66	\pm	0.02	$	& LasCumbres	\\
60575.29958	&	o & $ 	17.42	\pm	0.04	$	& ATLAS	\\
60575.33321	&	r & $ 	17.37	\pm	0.02	$	& ZTF	\\
60575.40036	&	g & $ 	17.08	\pm	0.02	$	& ZTF	\\
60575.40036	&	g & $ 	17.15	\pm	0.04	$	& ZTF	\\
60576.29561	&	i & $ 	17.49	\pm	0.03	$	& ZTF	\\
60576.33773	&	g & $ 	17.15	\pm	0.04	$	& ZTF	\\
60576.40149	&	r & $ 	17.39	\pm	0.01	$	& ZTF	\\
60577.31218	&	i & $ 	17.56	\pm	0.02	$	& ZTF	\\
60577.35797	&	g & $ 	17.11	\pm	0.04	$	& ZTF	\\
60577.4242	&	r & $ 	17.35	\pm	0.01	$	& ZTF	\\
60578.05086	&	UVW1 & $ 	18.49	\pm	0.09	$	& Swift	\\
60578.05157	&	U & $ 	17.29	\pm	0.08	$	& Swift	\\
60578.05228	&	B & $ 	17.1	\pm	0.09	$	& Swift	\\
60578.05563	&	UVW2 & $ 	19.66	\pm	0.12	$	& Swift	\\
60578.05634	&	V & $ 	17.22	\pm	0.17	$	& Swift	\\
60578.06043	&	UVM2 & $ 	19.19	\pm	0.11	$	& Swift	\\
60580.41517	&	B & $ 	17.19	\pm	0.02	$	& LasCumbres	\\
60580.41685	&	B & $ 	17.19	\pm	0.02	$	& LasCumbres	\\
60580.4187	&	V & $ 	17.21	\pm	0.02	$	& LasCumbres	\\
60580.42006	&	V & $ 	17.19	\pm	0.02	$	& LasCumbres	\\
60580.42158	&	g & $ 	17.04	\pm	0.01	$	& LasCumbres	\\
60580.42327	&	g & $ 	17.01	\pm	0.01	$	& LasCumbres	\\
60580.42513	&	r & $ 	17.34	\pm	0.02	$	& LasCumbres	\\
60580.42649	&	r & $ 	17.35	\pm	0.02	$	& LasCumbres	\\
60580.42804	&	i & $ 	17.56	\pm	0.02	$	& LasCumbres	\\
60580.42939	&	i & $ 	17.54	\pm	0.02	$	& LasCumbres	\\
60580.45544	&	r & $ 	17.29	\pm	0.01	$	& ZTF	\\
60580.46617	&	c & $ 	17.16	\pm	0.02	$	& ATLAS	\\
60580.50571	&	g & $ 	17.04	\pm	0.04	$	& ZTF	\\
60581.08823	&	o & $ 	17.23	\pm	0.12	$	& ATLAS	\\
60581.38796	&	UVW1 & $ 	18.36	\pm	0.11	$	& Swift	\\
60581.38846	&	U & $ 	17.22	\pm	0.09	$	& Swift	\\
60581.38895	&	B & $ 	17.06	\pm	0.11	$	& Swift	\\
60581.39122	&	UVW2 & $ 	19.4	\pm	0.13	$	& Swift	\\
60581.39515	&	UVM2 & $ 	18.95	\pm	0.1	$	& Swift	\\
60582.47732	&	o & $ 	17.4	\pm	0.02	$	& ATLAS	\\
60583.26336	&	c & $ 	17.07	\pm	0.01	$	& ATLAS	\\
60583.41552	&	r & $ 	17.27	\pm	0.01	$	& ZTF	\\
60583.4456	&	g & $ 	16.99	\pm	0.04	$	& ZTF	\\
60584.49219	&	c & $ 	17.11	\pm	0.01	$	& ATLAS	\\
60584.9204	&	o & $ 	17.3	\pm	0.02	$	& ATLAS	\\
60585.08574	&	B & $ 	17.11	\pm	0.02	$	& LasCumbres	\\
60585.08743	&	B & $ 	17.13	\pm	0.02	$	& LasCumbres	\\
60585.08928	&	V & $ 	17.09	\pm	0.02	$	& LasCumbres	\\
60585.09063	&	V & $ 	17.12	\pm	0.03	$	& LasCumbres	\\
60585.09216	&	g & $ 	16.89	\pm	0.01	$	& LasCumbres	\\
60585.09384	&	g & $ 	16.89	\pm	0.01	$	& LasCumbres	\\
60585.0957	&	r & $ 	17.22	\pm	0.02	$	& LasCumbres	\\
60585.09705	&	r & $ 	17.21	\pm	0.01	$	& LasCumbres	\\
60585.09858	&	i & $ 	17.38	\pm	0.03	$	& LasCumbres	\\
60585.09993	&	i & $ 	17.41	\pm	0.02	$	& LasCumbres	\\
60586.48205	&	o & $ 	17.34	\pm	0.02	$	& ATLAS	\\
60587.29042	&	c & $ 	17.03	\pm	0.01	$	& ATLAS	\\
60588.29078	&	B & $ 	17.08	\pm	0.02	$	& LasCumbres	\\
60588.29247	&	B & $ 	17.06	\pm	0.02	$	& LasCumbres	\\
60588.29431	&	V & $ 	17.05	\pm	0.02	$	& LasCumbres	\\
60588.29566	&	V & $ 	17.05	\pm	0.02	$	& LasCumbres	\\
60588.29721	&	g & $ 	16.79	\pm	0.01	$	& LasCumbres	\\
60588.2989	&	g & $ 	16.8	\pm	0.01	$	& LasCumbres	\\
60588.30074	&	r & $ 	17.25	\pm	0.02	$	& LasCumbres	\\
60588.30211	&	r & $ 	17.15	\pm	0.01	$	& LasCumbres	\\
60588.30365	&	i & $ 	17.37	\pm	0.01	$	& LasCumbres	\\
60588.30501	&	i & $ 	17.38	\pm	0.02	$	& LasCumbres	\\
60588.40855	&	c & $ 	17.01	\pm	0.01	$	& ATLAS	\\
60588.90388	&	o & $ 	17.3	\pm	0.02	$	& ATLAS	\\
60592.32478	&	r & $ 	17.2	\pm	0.01	$	& ZTF	\\
60592.37622	&	g & $ 	16.91	\pm	0.04	$	& ZTF	\\
60592.46155	&	c & $ 	17.01	\pm	0.01	$	& ATLAS	\\
60592.80457	&	o & $ 	17.27	\pm	0.03	$	& ATLAS	\\
60592.91576	&	B & $ 	17	\pm	0.02	$	& LasCumbres	\\
60592.91745	&	B & $ 	16.99	\pm	0.02	$	& LasCumbres	\\
60592.91937	&	V & $ 	17.03	\pm	0.02	$	& LasCumbres	\\
60592.92073	&	V & $ 	17.06	\pm	0.02	$	& LasCumbres	\\
60592.92229	&	g & $ 	16.82	\pm	0.01	$	& LasCumbres	\\
60592.92399	&	g & $ 	16.83	\pm	0.01	$	& LasCumbres	\\
60592.92589	&	r & $ 	17.17	\pm	0.01	$	& LasCumbres	\\
60592.92725	&	r & $ 	17.17	\pm	0.01	$	& LasCumbres	\\
60592.92882	&	i & $ 	17.37	\pm	0.02	$	& LasCumbres	\\
60592.93016	&	i & $ 	17.35	\pm	0.02	$	& LasCumbres	\\
60594.31713	&	g & $ 	16.89	\pm	0.04	$	& ZTF	\\
60594.38475	&	r & $ 	17.14	\pm	0.01	$	& ZTF	\\
60596.30442	&	g & $ 	16.9	\pm	0.04	$	& ZTF	\\
60596.37146	&	r & $ 	17.13	\pm	0.01	$	& ZTF	\\
60596.38325	&	g & $ 	16.79	\pm	0.05	$	& FLWO 1.2m	\\
60596.39518	&	r & $ 	17.2	\pm	0.07	$	& FLWO 1.2m	\\
60596.40696	&	i & $ 	17.35	\pm	0.07	$	& FLWO 1.2m	\\
60596.47553	&	o & $ 	17.19	\pm	0.01	$	& ATLAS	\\
60597.33115	&	g & $ 	16.78	\pm	0.02	$	& FLWO 1.2m	\\
60597.34293	&	r & $ 	17.2	\pm	0.1	$	& FLWO 1.2m	\\
60597.35465	&	i & $ 	17.39	\pm	0.08	$	& FLWO 1.2m	\\
60597.9236	&	B & $ 	17	\pm	0.02	$	& LasCumbres	\\
60597.92546	&	V & $ 	16.96	\pm	0.02	$	& LasCumbres	\\
60597.92699	&	g & $ 	16.76	\pm	0.01	$	& LasCumbres	\\
60597.92884	&	r & $ 	17.12	\pm	0.02	$	& LasCumbres	\\
60597.93035	&	i & $ 	17.4	\pm	0.03	$	& LasCumbres	\\
60598.24938	&	g & $ 	16.84	\pm	0.04	$	& ZTF	\\
60598.33844	&	r & $ 	17.13	\pm	0.03	$	& ZTF	\\
60601.91604	&	B & $ 	16.99	\pm	0.02	$	& LasCumbres	\\
60601.91788	&	V & $ 	16.98	\pm	0.02	$	& LasCumbres	\\
60601.9194	&	g & $ 	16.82	\pm	0.01	$	& LasCumbres	\\
60601.92125	&	r & $ 	17.17	\pm	0.02	$	& LasCumbres	\\
60601.92276	&	i & $ 	17.31	\pm	0.02	$	& LasCumbres	\\
60602.35936	&	r & $ 	17.14	\pm	0.06	$	& ZTF	\\
60602.38047	&	g & $ 	16.87	\pm	0.05	$	& ZTF	\\
60602.47506	&	o & $ 	17.2	\pm	0.03	$	& ATLAS	\\
60603.23955	&	o & $ 	17.22	\pm	0.03	$	& ATLAS	\\
60604.0666	&	UVW1 & $ 	18.3	\pm	0.1	$	& Swift	\\
60604.06709	&	U & $ 	17.15	\pm	0.09	$	& Swift	\\
60604.06758	&	B & $ 	16.89	\pm	0.1	$	& Swift	\\
60604.06986	&	UVW2 & $ 	19.49	\pm	0.14	$	& Swift	\\
60604.07035	&	V & $ 	16.98	\pm	0.18	$	& Swift	\\
60604.07362	&	UVM2 & $ 	18.97	\pm	0.11	$	& Swift	\\
60604.21892	&	r & $ 	17.02	\pm	0.01	$	& ZTF	\\
60604.26126	&	g & $ 	16.92	\pm	0.04	$	& ZTF	\\
60604.29792	&	B & $ 	17.07	\pm	0.02	$	& LasCumbres	\\
60604.29977	&	V & $ 	17.03	\pm	0.02	$	& LasCumbres	\\
60604.3013	&	g & $ 	16.87	\pm	0.02	$	& LasCumbres	\\
60604.30318	&	r & $ 	17.14	\pm	0.02	$	& LasCumbres	\\
60604.3047	&	i & $ 	17.31	\pm	0.02	$	& LasCumbres	\\
60604.41704	&	o & $ 	17.26	\pm	0.02	$	& ATLAS	\\
60606.24943	&	g & $ 	16.87	\pm	0.05	$	& ZTF	\\
60606.29428	&	r & $ 	17.15	\pm	0.01	$	& ZTF	\\
60607.17208	&	o & $ 	17.22	\pm	0.02	$	& ATLAS	\\
60607.59041	&	B & $ 	17.01	\pm	0.01	$	& LasCumbres	\\
60607.5923	&	V & $ 	17.03	\pm	0.02	$	& LasCumbres	\\
60607.59384	&	g & $ 	16.93	\pm	0.01	$	& LasCumbres	\\
60607.59573	&	r & $ 	17.18	\pm	0.01	$	& LasCumbres	\\
60607.59727	&	i & $ 	17.34	\pm	0.02	$	& LasCumbres	\\
60607.67055	&	UVW1 & $ 	18.32	\pm	0.08	$	& Swift	\\
60607.67161	&	U & $ 	17.16	\pm	0.06	$	& Swift	\\
60607.67266	&	B & $ 	16.93	\pm	0.07	$	& Swift	\\
60607.67776	&	UVW2 & $ 	19.35	\pm	0.09	$	& Swift	\\
60607.67882	&	V & $ 	16.89	\pm	0.11	$	& Swift	\\
60607.68612	&	UVM2 & $ 	19	\pm	0.08	$	& Swift	\\
60608.27549	&	g & $ 	16.88	\pm	0.05	$	& ZTF	\\
60608.38215	&	r & $ 	17.04	\pm	0.01	$	& ZTF	\\
60608.44519	&	o & $ 	17.14	\pm	0.13	$	& ATLAS	\\
60608.94	&	c & $ 	16.96	\pm	0.01	$	& ATLAS	\\
60610.09367	&	B & $ 	17.02	\pm	0.02	$	& LasCumbres	\\
60610.09551	&	V & $ 	17	\pm	0.02	$	& LasCumbres	\\
60610.09705	&	g & $ 	16.77	\pm	0.01	$	& LasCumbres	\\
60610.09891	&	r & $ 	17.1	\pm	0.01	$	& LasCumbres	\\
60610.10043	&	i & $ 	17.24	\pm	0.02	$	& LasCumbres	\\
60610.20253	&	r & $ 	17.08	\pm	0.08	$	& ZTF	\\
60610.26354	&	g & $ 	16.9	\pm	0.06	$	& ZTF	\\
60611.18551	&	o & $ 	17.05	\pm	0.02	$	& ATLAS	\\
60612.51253	&	o & $ 	17.28	\pm	0.04	$	& ATLAS	\\
60613.32528	&	B & $ 	17.14	\pm	0.02	$	& LasCumbres	\\
60613.32718	&	V & $ 	17.06	\pm	0.02	$	& LasCumbres	\\
60613.32869	&	g & $ 	16.9	\pm	0.01	$	& LasCumbres	\\
60613.33061	&	r & $ 	17.17	\pm	0.01	$	& LasCumbres	\\
60613.33211	&	i & $ 	17.32	\pm	0.02	$	& LasCumbres	\\
60614.38181	&	g & $ 	16.91	\pm	0.05	$	& ZTF	\\
60614.40798	&	c & $ 	17.01	\pm	0.01	$	& ATLAS	\\
60614.42259	&	r & $ 	17.14	\pm	0.01	$	& ZTF	\\
60615.15098	&	o & $ 	17.16	\pm	0.01	$	& ATLAS	\\
60615.85226	&	B & $ 	17.06	\pm	0.02	$	& LasCumbres	\\
60615.8541	&	V & $ 	17.05	\pm	0.02	$	& LasCumbres	\\
60615.85561	&	g & $ 	16.83	\pm	0.01	$	& LasCumbres	\\
60615.85745	&	r & $ 	17.13	\pm	0.01	$	& LasCumbres	\\
60615.85897	&	i & $ 	17.25	\pm	0.02	$	& LasCumbres	\\
60616.33112	&	r & $ 	17.15	\pm	0.01	$	& ZTF	\\
60616.3594	&	g & $ 	16.9	\pm	0.05	$	& ZTF	\\
60616.39742	&	o & $ 	17.17	\pm	0.01	$	& ATLAS	\\
60616.96868	&	c & $ 	17.04	\pm	0.1	$	& ATLAS	\\
60618.37479	&	o & $ 	17.23	\pm	0.02	$	& ATLAS	\\
60619.07703	&	B & $ 	17.03	\pm	0.01	$	& LasCumbres	\\
60619.07891	&	V & $ 	17.02	\pm	0.02	$	& LasCumbres	\\
60619.08045	&	g & $ 	16.8	\pm	0.01	$	& LasCumbres	\\
60619.08232	&	r & $ 	17.16	\pm	0.01	$	& LasCumbres	\\
60619.08383	&	i & $ 	17.28	\pm	0.01	$	& LasCumbres	\\
60619.14068	&	o & $ 	17.2	\pm	0.02	$	& ATLAS	\\
60619.16567	&	r & $ 	17.16	\pm	0.01	$	& ZTF	\\
60619.3172	&	g & $ 	16.9	\pm	0.04	$	& ZTF	\\
60620.90475	&	o & $ 	17.22	\pm	0.02	$	& ATLAS	\\
60621.17069	&	g & $ 	16.9	\pm	0.04	$	& ZTF	\\
60621.19421	&	r & $ 	17.17	\pm	0.03	$	& ZTF	\\
60621.82406	&	B & $ 	17	\pm	0.03	$	& LasCumbres	\\
60621.82595	&	V & $ 	17.01	\pm	0.03	$	& LasCumbres	\\
60621.82748	&	g & $ 	16.82	\pm	0.02	$	& LasCumbres	\\
60621.82934	&	r & $ 	17.2	\pm	0.04	$	& LasCumbres	\\
60621.83087	&	i & $ 	17.2	\pm	0.05	$	& LasCumbres	\\
60622.18421	&	r & $ 	17.18	\pm	0.01	$	& ZTF	\\
60622.19249	&	g & $ 	16.92	\pm	0.04	$	& ZTF	\\
60623.14738	&	o & $ 	17.15	\pm	0.02	$	& ATLAS	\\
60624.29586	&	r & $ 	17.11	\pm	0.02	$	& ZTF	\\
60624.34058	&	g & $ 	16.93	\pm	0.03	$	& ZTF	\\
60624.53505	&	o & $ 	17.2	\pm	0.02	$	& ATLAS	\\
60625.11411	&	B & $ 	17.06	\pm	0.02	$	& LasCumbres	\\
60625.11599	&	V & $ 	17.06	\pm	0.02	$	& LasCumbres	\\
60625.11753	&	g & $ 	16.83	\pm	0.01	$	& LasCumbres	\\
60625.11939	&	r & $ 	17.18	\pm	0.01	$	& LasCumbres	\\
60625.1209	&	i & $ 	17.33	\pm	0.02	$	& LasCumbres	\\
60625.84348	&	B & $ 	17.05	\pm	0.03	$	& LasCumbres	\\
60625.84534	&	V & $ 	17.03	\pm	0.04	$	& LasCumbres	\\
60625.84685	&	g & $ 	16.86	\pm	0.02	$	& LasCumbres	\\
60625.8487	&	r & $ 	17.25	\pm	0.03	$	& LasCumbres	\\
60625.85022	&	i & $ 	17.38	\pm	0.03	$	& LasCumbres	\\
60626.18683	&	r & $ 	17.18	\pm	0.04	$	& ZTF	\\
60626.21508	&	g & $ 	16.93	\pm	0.04	$	& ZTF	\\
60628.85356	&	B & $ 	17.08	\pm	0.02	$	& LasCumbres	\\
60628.85543	&	V & $ 	17.08	\pm	0.02	$	& LasCumbres	\\
60628.85694	&	g & $ 	16.86	\pm	0.01	$	& LasCumbres	\\
60628.85881	&	r & $ 	17.2	\pm	0.02	$	& LasCumbres	\\
60628.86035	&	i & $ 	17.37	\pm	0.03	$	& LasCumbres	\\
60630.46921	&	o & $ 	17.39	\pm	0.05	$	& ATLAS	\\
60631.1549	&	o & $ 	17.3	\pm	0.04	$	& ATLAS	\\
60631.86398	&	B & $ 	17.12	\pm	0.02	$	& LasCumbres	\\
60631.86581	&	V & $ 	17.09	\pm	0.04	$	& LasCumbres	\\
60631.86732	&	g & $ 	16.92	\pm	0.01	$	& LasCumbres	\\
60631.86916	&	r & $ 	17.27	\pm	0.01	$	& LasCumbres	\\
60631.87071	&	i & $ 	17.35	\pm	0.02	$	& LasCumbres	\\
60632.01548	&	UVW1 & $ 	18.86	\pm	0.17	$	& Swift	\\
60632.01592	&	U & $ 	17.38	\pm	0.11	$	& Swift	\\
60632.01637	&	B & $ 	17.19	\pm	0.13	$	& Swift	\\
60632.01842	&	UVW2 & $ 	19.77	\pm	0.2	$	& Swift	\\
60632.01887	&	V & $ 	17.03	\pm	0.2	$	& Swift	\\
60632.02154	&	UVM2 & $ 	19.22	\pm	0.15	$	& Swift	\\
60632.22647	&	r & $ 	17.16	\pm	0.01	$	& ZTF	\\
60632.29935	&	g & $ 	17	\pm	0.06	$	& ZTF	\\
60632.87592	&	o & $ 	17.28	\pm	0.02	$	& ATLAS	\\
60634.20856	&	r & $ 	17.31	\pm	0.03	$	& ZTF	\\
60634.26944	&	g & $ 	17.01	\pm	0.04	$	& ZTF	\\
60634.35721	&	o & $ 	17.32	\pm	0.02	$	& ATLAS	\\
60635.1213	&	c & $ 	17.08	\pm	0.01	$	& ATLAS	\\
60635.17431	&	r & $ 	17.17	\pm	0.01	$	& ZTF	\\
60635.29994	&	g & $ 	17.02	\pm	0.05	$	& ZTF	\\
60635.41425	&	B & $ 	17.23	\pm	0.02	$	& LasCumbres	\\
60635.41612	&	V & $ 	17.12	\pm	0.02	$	& LasCumbres	\\
60635.4177	&	g & $ 	16.97	\pm	0.01	$	& LasCumbres	\\
60635.41956	&	r & $ 	17.23	\pm	0.01	$	& LasCumbres	\\
60635.42111	&	i & $ 	17.34	\pm	0.01	$	& LasCumbres	\\
60636.35558	&	c & $ 	17.11	\pm	0.01	$	& ATLAS	\\
60636.74494	&	UVW1 & $ 	18.75	\pm	0.09	$	& Swift	\\
60636.74601	&	U & $ 	17.47	\pm	0.07	$	& Swift	\\
60636.74709	&	B & $ 	17	\pm	0.07	$	& Swift	\\
60636.74793	&	UVW2 & $ 	19.66	\pm	0.24	$	& Swift	\\
60636.84146	&	o & $ 	17.32	\pm	0.02	$	& ATLAS	\\
60637.20602	&	r & $ 	17.18	\pm	0.01	$	& ZTF	\\
60637.25483	&	g & $ 	17.03	\pm	0.04	$	& ZTF	\\
60637.71403	&	UVW1 & $ 	18.83	\pm	0.1	$	& Swift	\\
60637.7151	&	U & $ 	17.42	\pm	0.07	$	& Swift	\\
60637.71619	&	B & $ 	17.06	\pm	0.08	$	& Swift	\\
60637.72134	&	UVW2 & $ 	19.96	\pm	0.12	$	& Swift	\\
60637.72242	&	V & $ 	17.05	\pm	0.12	$	& Swift	\\
60637.72848	&	UVM2 & $ 	19.45	\pm	0.1	$	& Swift	\\
60638.35067	&	o & $ 	17.34	\pm	0.01	$	& ATLAS	\\
60639.10152	&	c & $ 	17.08	\pm	0.01	$	& ATLAS	\\
60639.54364	&	B & $ 	17.25	\pm	0.02	$	& LasCumbres	\\
60639.54552	&	V & $ 	17.14	\pm	0.02	$	& LasCumbres	\\
60639.54707	&	g & $ 	17.01	\pm	0.01	$	& LasCumbres	\\
60639.54894	&	r & $ 	17.29	\pm	0.01	$	& LasCumbres	\\
60639.55047	&	i & $ 	17.37	\pm	0.02	$	& LasCumbres	\\
60640.1525	&	r & $ 	17.22	\pm	0.01	$	& ZTF	\\
60640.19132	&	g & $ 	17.06	\pm	0.04	$	& ZTF	\\
60640.19132	&	g & $ 	17.06	\pm	0.04	$	& ZTF	\\
60640.34096	&	c & $ 	17.13	\pm	0.02	$	& ATLAS	\\
60642.69903	&	UVW1 & $ 	18.93	\pm	0.16	$	& Swift	\\
60642.69941	&	U & $ 	17.37	\pm	0.11	$	& Swift	\\
60642.69979	&	B & $ 	17.21	\pm	0.13	$	& Swift	\\
60642.70151	&	UVW2 & $ 	20.06	\pm	0.2	$	& Swift	\\
60642.70189	&	V & $ 	17.21	\pm	0.23	$	& Swift	\\
60642.70418	&	UVM2 & $ 	19.61	\pm	0.17	$	& Swift	\\
60643.09428	&	c & $ 	17.1	\pm	0.02	$	& ATLAS	\\
60643.0974	&	B & $ 	17.23	\pm	0.02	$	& LasCumbres	\\
60643.09927	&	V & $ 	17.16	\pm	0.02	$	& LasCumbres	\\
60643.10082	&	g & $ 	16.97	\pm	0.01	$	& LasCumbres	\\
60643.10267	&	r & $ 	17.28	\pm	0.01	$	& LasCumbres	\\
60643.10418	&	i & $ 	17.35	\pm	0.02	$	& LasCumbres	\\
60644.28506	&	g & $ 	17.09	\pm	0.03	$	& ZTF	\\
60644.31747	&	c & $ 	17.12	\pm	0.03	$	& ATLAS	\\
60644.82192	&	o & $ 	17.35	\pm	0.02	$	& ATLAS	\\
60646.13518	&	B & $ 	17.24	\pm	0.02	$	& LasCumbres	\\
60646.13703	&	V & $ 	17.14	\pm	0.02	$	& LasCumbres	\\
60646.13853	&	g & $ 	17.02	\pm	0.01	$	& LasCumbres	\\
60646.14039	&	r & $ 	17.27	\pm	0.01	$	& LasCumbres	\\
60646.1419	&	i & $ 	17.4	\pm	0.02	$	& LasCumbres	\\
60647.08383	&	c & $ 	17.13	\pm	0.01	$	& ATLAS	\\
60647.18697	&	r & $ 	17.23	\pm	0.01	$	& ZTF	\\
60647.21941	&	g & $ 	17.08	\pm	0.04	$	& ZTF	\\
60648.81423	&	o & $ 	17.34	\pm	0.03	$	& ATLAS	\\
60649.1138	&	B & $ 	17.28	\pm	0.02	$	& LasCumbres	\\
60649.11566	&	V & $ 	17.15	\pm	0.02	$	& LasCumbres	\\
60649.11718	&	g & $ 	17.04	\pm	0.01	$	& LasCumbres	\\
60649.11903	&	r & $ 	17.26	\pm	0.01	$	& LasCumbres	\\
60649.12053	&	i & $ 	17.35	\pm	0.02	$	& LasCumbres	\\
60649.2074	&	g & $ 	17.05	\pm	0.04	$	& ZTF	\\
60649.24772	&	r & $ 	17.24	\pm	0.01	$	& ZTF	\\
60650.30072	&	o & $ 	17.35	\pm	0.02	$	& ATLAS	\\
60651.08776	&	o & $ 	17.18	\pm	0.02	$	& ATLAS	\\
60651.14631	&	r & $ 	17.22	\pm	0.01	$	& ZTF	\\
60651.21161	&	g & $ 	17.04	\pm	0.04	$	& ZTF	\\
60652.10808	&	UVW1 & $ 	18.95	\pm	0.1	$	& Swift	\\
60652.1091	&	U & $ 	17.36	\pm	0.07	$	& Swift	\\
60652.11011	&	B & $ 	16.98	\pm	0.07	$	& Swift	\\
60652.11212	&	g & $ 	17	\pm	0.1	$	& FLWO 1.2m	\\
60652.11346	&	B & $ 	17.22	\pm	0.02	$	& LasCumbres	\\
60652.11499	&	UVW2 & $ 	19.96	\pm	0.12	$	& Swift	\\
60652.11532	&	V & $ 	17.13	\pm	0.02	$	& LasCumbres	\\
60652.116	&	V & $ 	17.17	\pm	0.14	$	& Swift	\\
60652.11688	&	g & $ 	17	\pm	0.01	$	& LasCumbres	\\
60652.11873	&	r & $ 	17.3	\pm	0.01	$	& LasCumbres	\\
60652.12024	&	i & $ 	17.36	\pm	0.02	$	& LasCumbres	\\
60652.12085	&	UVM2 & $ 	19.64	\pm	0.12	$	& Swift	\\
60652.12717	&	r & $ 	17.39	\pm	0.09	$	& FLWO 1.2m	\\
60652.13855	&	i & $ 	17.41	\pm	0.08	$	& FLWO 1.2m	\\
60652.56798	&	o & $ 	17.3	\pm	0.02	$	& ATLAS	\\
60653.10991	&	r & $ 	17.21	\pm	0.02	$	& ZTF	\\
60653.19316	&	g & $ 	17.03	\pm	0.04	$	& ZTF	\\
60656.17093	&	B & $ 	17.25	\pm	0.03	$	& LasCumbres	\\
60656.1729	&	V & $ 	17.12	\pm	0.03	$	& LasCumbres	\\
60656.17451	&	g & $ 	17	\pm	0.02	$	& LasCumbres	\\
60656.17637	&	r & $ 	17.26	\pm	0.05	$	& LasCumbres	\\
60656.17794	&	i & $ 	17.35	\pm	0.07	$	& LasCumbres	\\
60657.92689	&	UVW1 & $ 	18.71	\pm	0.13	$	& Swift	\\
60657.92734	&	U & $ 	17.5	\pm	0.1	$	& Swift	\\
60657.92779	&	B & $ 	16.98	\pm	0.11	$	& Swift	\\
60657.92988	&	UVW2 & $ 	19.63	\pm	0.15	$	& Swift	\\
60657.93033	&	V & $ 	17.24	\pm	0.22	$	& Swift	\\
60658.17101	&	r & $ 	17.22	\pm	0.02	$	& ZTF	\\
60658.22712	&	g & $ 	17.07	\pm	0.05	$	& ZTF	\\
60658.35288	&	o & $ 	17.41	\pm	0.04	$	& ATLAS	\\
60659.10318	&	o & $ 	17.31	\pm	0.05	$	& ATLAS	\\
60659.15252	&	B & $ 	17.22	\pm	0.02	$	& LasCumbres	\\
60659.15436	&	V & $ 	17.06	\pm	0.02	$	& LasCumbres	\\
60659.15587	&	g & $ 	17.02	\pm	0.01	$	& LasCumbres	\\
60659.15773	&	r & $ 	17.32	\pm	0.02	$	& LasCumbres	\\
60659.15925	&	i & $ 	17.34	\pm	0.02	$	& LasCumbres	\\
60660.1865	&	r & $ 	17.29	\pm	0.04	$	& ZTF	\\
60660.5793	&	o & $ 	17.36	\pm	0.02	$	& ATLAS	\\
60662.21107	&	B & $ 	17.29	\pm	0.02	$	& LasCumbres	\\
60662.21292	&	V & $ 	17.18	\pm	0.02	$	& LasCumbres	\\
60662.21445	&	g & $ 	17.05	\pm	0.01	$	& LasCumbres	\\
60662.2163	&	r & $ 	17.36	\pm	0.02	$	& LasCumbres	\\
60662.21781	&	i & $ 	17.38	\pm	0.02	$	& LasCumbres	\\
60663.07996	&	o & $ 	17.34	\pm	0.02	$	& ATLAS	\\
60663.09932	&	UVW1 & $ 	18.84	\pm	0.16	$	& Swift	\\
60663.09967	&	U & $ 	17.59	\pm	0.13	$	& Swift	\\
60663.10002	&	B & $ 	17.25	\pm	0.15	$	& Swift	\\
60663.10159	&	UVW2 & $ 	19.72	\pm	0.18	$	& Swift	\\
60663.10418	&	UVM2 & $ 	19.46	\pm	0.16	$	& Swift	\\
60663.23331	&	g & $ 	17.14	\pm	0.05	$	& ZTF	\\
60663.27782	&	r & $ 	17.38	\pm	0.02	$	& ZTF	\\
60664.32328	&	o & $ 	17.37	\pm	0.02	$	& ATLAS	\\
60664.89897	&	c & $ 	17.22	\pm	0.02	$	& ATLAS	\\
60665.08285	&	B & $ 	17.37	\pm	0.02	$	& LasCumbres	\\
60665.08472	&	V & $ 	17.19	\pm	0.02	$	& LasCumbres	\\
60665.08624	&	g & $ 	17.04	\pm	0.01	$	& LasCumbres	\\
60665.08809	&	r & $ 	17.31	\pm	0.01	$	& LasCumbres	\\
60665.0896	&	i & $ 	17.4	\pm	0.02	$	& LasCumbres	\\
60665.1497	&	r & $ 	17.3	\pm	0.01	$	& ZTF	\\
60666.29883	&	o & $ 	17.39	\pm	0.01	$	& ATLAS	\\
60666.42093	&	UVW1 & $ 	18.88	\pm	0.1	$	& Swift	\\
60666.42199	&	U & $ 	17.62	\pm	0.07	$	& Swift	\\
60666.42305	&	B & $ 	17.15	\pm	0.08	$	& Swift	\\
60666.42813	&	UVW2 & $ 	19.94	\pm	0.12	$	& Swift	\\
60666.42919	&	V & $ 	17.06	\pm	0.13	$	& Swift	\\
60666.43682	&	UVM2 & $ 	19.56	\pm	0.1	$	& Swift	\\
60667.04482	&	o & $ 	17.22	\pm	0.04	$	& ATLAS	\\
60667.12512	&	g & $ 	17.17	\pm	0.01	$	& ZTF	\\
60667.12512	&	g & $ 	17.2	\pm	0.04	$	& ZTF	\\
60667.17104	&	r & $ 	17.39	\pm	0.01	$	& ZTF	\\
60668.27983	&	o & $ 	17.44	\pm	0.02	$	& ATLAS	\\
60671.05392	&	UVW1 & $ 	19.03	\pm	0.17	$	& Swift	\\
60671.0546	&	U & $ 	17.89	\pm	0.13	$	& Swift	\\
60671.05529	&	B & $ 	17.54	\pm	0.16	$	& Swift	\\
60671.05847	&	o & $ 	17.53	\pm	0.03	$	& ATLAS	\\
60671.05853	&	UVW2 & $ >	20.16			$	& Swift	\\
60671.0639	&	UVM2 & $ 	19.95	\pm	0.18	$	& Swift	\\
60671.17083	&	r & $ 	17.58	\pm	0.08	$	& ZTF	\\
60672.30322	&	o & $ 	17.66	\pm	0.02	$	& ATLAS	\\
60672.79861	&	c & $ 	17.51	\pm	0.03	$	& ATLAS	\\
60673.14581	&	r & $ 	17.64	\pm	0.02	$	& ZTF	\\
60673.14581	&	r & $ 	17.62	\pm	0.06	$	& ZTF	\\
60674.26277	&	o & $ 	17.76	\pm	0.02	$	& ATLAS	\\
60675.06888	&	B & $ 	17.85	\pm	0.02	$	& LasCumbres	\\
60675.07072	&	V & $ 	17.6	\pm	0.02	$	& LasCumbres	\\
60675.07223	&	g & $ 	17.49	\pm	0.01	$	& LasCumbres	\\
60675.07408	&	r & $ 	17.66	\pm	0.02	$	& LasCumbres	\\
60675.07558	&	i & $ 	17.72	\pm	0.02	$	& LasCumbres	\\
60675.16513	&	r & $ 	17.69	\pm	0.02	$	& ZTF	\\
60675.21575	&	g & $ 	17.6	\pm	0.02	$	& ZTF	\\
60675.21575	&	g & $ 	17.62	\pm	0.04	$	& ZTF	\\
60676.16456	&	UVW1 & $ >	19.32			$	& Swift	\\
60676.16496	&	U & $ 	17.93	\pm	0.15	$	& Swift	\\
60676.16535	&	B & $ 	17.74	\pm	0.2	$	& Swift	\\
60676.16714	&	UVW2 & $ 	20.28	\pm	0.23	$	& Swift	\\
60676.16753	&	V & $ >	17.04			$	& Swift	\\
60676.17015	&	UVM2 & $ 	20.33	\pm	0.24	$	& Swift	\\
60677.15262	&	r & $ 	17.68	\pm	0.02	$	& ZTF	\\
60677.20718	&	g & $ 	17.69	\pm	0.02	$	& ZTF	\\
60677.20718	&	g & $ 	17.71	\pm	0.07	$	& ZTF	\\
60678.05177	&	B & $ 	18.02	\pm	0.02	$	& LasCumbres	\\
60678.05361	&	V & $ 	17.74	\pm	0.02	$	& LasCumbres	\\
60678.05512	&	g & $ 	17.66	\pm	0.01	$	& LasCumbres	\\
60678.05696	&	r & $ 	17.76	\pm	0.01	$	& LasCumbres	\\
60678.05846	&	i & $ 	17.83	\pm	0.02	$	& LasCumbres	\\
60678.25791	&	o & $ 	17.9	\pm	0.03	$	& ATLAS	\\
60679.04288	&	o & $ 	17.91	\pm	0.05	$	& ATLAS	\\
60679.15253	&	r & $ 	17.85	\pm	0.02	$	& ZTF	\\
60679.20623	&	g & $ 	17.88	\pm	0.02	$	& ZTF	\\
60679.20623	&	g & $ 	17.83	\pm	0.06	$	& ZTF	\\
60680.08632	&	r & $ 	17.98	\pm	0.07	$	& FLWO 1.2m	\\
60680.09787	&	i & $ 	17.95	\pm	0.08	$	& FLWO 1.2m	\\
60680.19827	&	UVW1 & $ >	18.37			$	& Swift	\\
60680.19857	&	U & $ 	18.49	\pm	0.33	$	& Swift	\\
60680.19888	&	B & $ >	17.25			$	& Swift	\\
60680.20023	&	UVW2 & $ >	19.29			$	& Swift	\\
60680.20053	&	V & $ >	16.55			$	& Swift	\\
60680.20279	&	UVM2 & $ >	19.39			$	& Swift	\\
60680.25254	&	o & $ 	17.91	\pm	0.03	$	& ATLAS	\\
60684.19609	&	r & $ 	17.9	\pm	0.1	$	& ZTF	\\
60684.21501	&	g & $ 	18.09	\pm	0.07	$	& ZTF	\\
60684.21501	&	g & $ 	18.04	\pm	0.12	$	& ZTF	\\
60684.26064	&	o & $ 	17.97	\pm	0.04	$	& ATLAS	\\
60685.17112	&	UVW1 & $ 	19.95	\pm	0.2	$	& Swift	\\
60685.17218	&	U & $ 	18.63	\pm	0.14	$	& Swift	\\
60685.17239	&	B & $ >	17.2			$	& Swift	\\
60685.24548	&	UVW2 & $ >	20.63			$	& Swift	\\
60685.24615	&	V & $ >	17.37			$	& Swift	\\
60685.25071	&	UVM2 & $ 	21.16	\pm	0.29	$	& Swift	\\
60686.03727	&	B & $ 	18.4	\pm	0.04	$	& LasCumbres	\\
60686.03912	&	V & $ 	17.99	\pm	0.03	$	& LasCumbres	\\
60686.04064	&	g & $ 	18.04	\pm	0.03	$	& LasCumbres	\\
60686.0425	&	r & $ 	18.01	\pm	0.03	$	& LasCumbres	\\
60686.04402	&	i & $ 	17.99	\pm	0.03	$	& LasCumbres	\\
60686.13102	&	g & $ 	18.1	\pm	0.07	$	& ZTF	\\
60687.20652	&	r & $ 	18.2	\pm	0.1	$	& FLWO 1.2m	\\
60687.21835	&	i & $ 	18.1	\pm	0.06	$	& FLWO 1.2m	\\
60688.12924	&	r & $ 	18.11	\pm	0.04	$	& ZTF	\\
60688.18767	&	g & $ 	18.3	\pm	0.09	$	& ZTF	\\
60688.18767	&	g & $ 	18.23	\pm	0.12	$	& ZTF	\\
60688.79872	&	o & $ 	18.07	\pm	0.09	$	& ATLAS	\\
60690.21594	&	g & $ 	18.07	\pm	0.22	$	& ZTF	\\
60690.31665	&	o & $ 	18.08	\pm	0.06	$	& ATLAS	\\
60690.39256	&	UVW1 & $ >	19.64			$	& Swift	\\
60690.39329	&	U & $ 	19.29	\pm	0.33	$	& Swift	\\
60690.39402	&	B & $ >	18.07			$	& Swift	\\
60690.39748	&	UVW2 & $ >	20.4			$	& Swift	\\
60690.39821	&	V & $ >	17.4			$	& Swift	\\
60690.40348	&	UVM2 & $ 	20.84	\pm	0.29	$	& Swift	\\
60692.80117	&	o & $ 	18.13	\pm	0.07	$	& ATLAS	\\
60694.12317	&	r & $ 	18.29	\pm	0.04	$	& ZTF	\\
60694.12317	&	r & $ 	18.24	\pm	0.08	$	& ZTF	\\
60694.14484	&	g & $ 	18.32	\pm	0.04	$	& ZTF	\\
60694.14484	&	g & $ 	18.26	\pm	0.08	$	& ZTF	\\
60694.27096	&	o & $ 	18.24	\pm	0.05	$	& ATLAS	\\
60696.25128	&	c & $ 	18.27	\pm	0.03	$	& ATLAS	\\
60699.15138	&	g & $ 	18.5	\pm	0.06	$	& ZTF	\\
60699.15138	&	g & $ 	18.43	\pm	0.12	$	& ZTF	\\
60706.13259	&	g & $ 	18.53	\pm	0.04	$	& ZTF	\\
60706.13259	&	g & $ 	18.61	\pm	0.09	$	& ZTF	\\
60711.08773	&	B & $ 	19.14	\pm	0.07	$	& LasCumbres	\\
60711.09075	&	V & $ 	18.52	\pm	0.04	$	& LasCumbres	\\
60711.09259	&	g & $ 	18.63	\pm	0.03	$	& LasCumbres	\\
60711.09538	&	r & $ 	18.54	\pm	0.03	$	& LasCumbres	\\
60711.09722	&	i & $ 	18.44	\pm	0.05	$	& LasCumbres	\\
60711.14999	&	r & $ 	18.41	\pm	0.11	$	& ZTF	\\

\hline

\caption{Photometry of SN\,2024rmj. Magnitudes are in the AB system and corrected for Galactic Extinction in the direction of SN\,2024rmj.}
\label{tab:photometrytable}
\end{longtable}

\bibliography{ref.bib}{}
\bibliographystyle{aasjournal}

\end{document}